\documentclass[pre,aps,floats,twocolumn,superscriptaddress,floatfix]{revtex4}
\usepackage{graphicx,amssymb,amsmath,ifthen,rotating,chemfig}
\usepackage[active]{srcltx}
\begin{document}
\title{Free-energy landscapes and insertion pathways for peptides in membrane environment}  
\author{Ganga P. Sharma}
\affiliation{
  Department of Physics,
  University of Rhode Island,
  Kingston RI 02881, USA}
  \author{Aaron C. Meyer}
\affiliation{
  Department of Physics,
  University of Rhode Island,
  Kingston RI 02881, USA}
   \author{Suhail Habeeb}
\affiliation{
  Department of Physics,
  University of Rhode Island,
  Kingston RI 02881, USA}
        \author{Michael Karbach}
\affiliation{
Fachgruppe Physik,
  Bergische Universit{\"{a}}t Wuppertal,
  D-42097 Wuppertal, Germany}
\author{Gerhard M{\"{u}}ller}
\affiliation{
  Department of Physics,
  University of Rhode Island,
  Kingston RI 02881, USA}

\begin{abstract}
Free-energy landscapes for short peptides -- specifically for variants of the pH Low Insertion Peptide (pHLIP) -- in the heterogeneous environment of a lipid bilayer or cell membrane are constructed, taking into account a set of dominant interactions and the conformational preferences of the peptide backbone.
Our methodology interprets broken internal H-bonds along the backbone of a polypeptide as statistically interacting quasiparticles, activated from the helix reference state.
The favored conformation depends on the local environment (ranging from polar to nonpolar), specifically on the availability of external H-bonds (with $\mathrm{H_2O}$ molecules or lipid headgroups) to replace internal H-bonds.
The dominant side-chain contribution is accounted for by residue-specific transfer free energies between polar and nonpolar environments.
The free-energy landscape is sensitive to the level of pH in the aqueous environment surrounding the membrane. 
For high pH, we identify pathways of descending free energy that suggest a coexistence of membrane-adsorbed peptides with peptides in solution.
A drop in pH raises the degree of protonation of negatively charged residues and thus increases the hydrophobicity of peptide segments near the C terminus.
For low pH, we identify insertion pathways between the membrane-adsorbed state and a stable trans-membrane state with the C terminus having crossed the membrane.
\end{abstract}

\maketitle

%
\section{Introduction}\label{sec:intro}
%
Water-soluble peptides with an affinity to insertion into cell membranes under specific conditions have found applications in medical research as diagnostic and therapeutic tools. 
They have been shown to carry fluorescent markers or drugs to targeted regions in living organisms with a continually advancing degree of efficiency and discrimination \cite{AER10, WMT+13, YDW+13, SAS+13, AER14, Resh15, AME+16, AMG+16, RMAE20}.
Yet many open questions remain and call for answers.
Analytic studies and computational studies tend to be complementary in their strengths and limitations to provide answers.
Both approaches are needed to advance our knowledge of peptide insertion into lipid membranes.
Molecular dynamics studies offer high resolution at short time scales, whereas kinetic studies, which combine analytic and computational aspects in different ways, are better equipped for dealing with processes that involve multiple time scales.

This work reports the second stage of a three-stage project aiming to support a more complete theoretical understanding of membrane-associated protein or peptide folding \cite{Jaeh83, BBH96, BBNH96, KB02, RGHM04, Bowi05, HKB+05, OSMG15, VCC+04, ORV05, WV04, SSDI10, BYMB06, Rent10}.
The first stage, reported in Ref.~\cite{cohetra}, involved the design and solution of a microscopic model for the coil-helix transition of a long polypeptide adsorbed to a planar water-lipid interface.
The methodology interprets the polypeptide and its homogeneous, effectively two-dimensional environment as a system of statistically interacting quasiparticles activated from the (ordered) helix state.
The particles represent broken internal H-bonds along the backbone, producing segments of a (disordered) coil conformation, sprawled across the interface in the shape of a self-avoiding random walk.

In one experimentally realized scenario \cite{RSA+07, ADS+07, RAS+08, AKW+10, KWW+12, WMT+13, WAR16, VSS+18, JW89, ES81, PE90, ECC+03, RMAE20}, the coil-helix conversion is triggered by a drop in pH.
The ensuing protonation of negatively charged side chains enhances the hydrophobicity of the polypeptide and pushes its backbone deeper into the (nonpolar) membrane.
This environmental change favors the formation of internal hydrogen bonds, which stabilize the $\alpha$-helix conformation.
Depending on the parameter settings, the model predicts the conformation to change as a crossover, a first-order transition, or a second-order transition. 

Here, in the continuation of this project, we begin by considering long polypeptides which are no longer confined to a flat water-lipid interface, but are positioned and oriented along some line in the heterogeneous environment comprising the lipid bilayer of a liposome or a biological cell and the surrounding water.
The model parameter identified in Ref.~\cite{cohetra} to drive the conformational change then turns into a (scalar) field with values reflecting the nature of the local medium, specifically its degree of polarity.
Such circumstances pose a challenge to any method of analysis.
The methodology used here has already proven to be adaptable to heterogeneous environments in different applications \cite{NA14, janac2, inharo}.

From the analysis thus carried out, profiles emerge for the densities of free energy, enthalpy, entropy, and helicity pertaining to segments of long polypeptides in the heterogeneous membrane environment.
The profiles reflect enthalpic and entropic consequences of the interactions between the backbone of the polypeptide and the lipid membrane or the surrounding hydrogen-bonded network of water molecules.
These profiles are then taken to represent propensities for the statistical mechanical behavior of short peptides in the same environment.
In this step, enthalpic and entropic effects involving the side chains and the semi-fluid bilayer of lipid amphiphiles can be built into the model.
The result are landscapes of free energy, enthalpy, entropy, and helicity for short peptides of given composition.

The free-energy landscapes incorporate road signs for pathways that guide the peptide from solution to adsorption under one set of environmental circumstances and from adsorption to insertion under a different set.
The enthalpy and entropy landscapes offer insights into the dominant forces that shape the pathways of descending free energy.
This analysis, which is grounded in equilibrium statistical mechanics, will be a key ingredient for the third stage of this project \cite{kinpep}: a kinetics study of trans-membrane insertion and exit of peptides.

\begin{figure}[htb]
  \begin{center}
  \includegraphics[width=23mm,angle=-90]{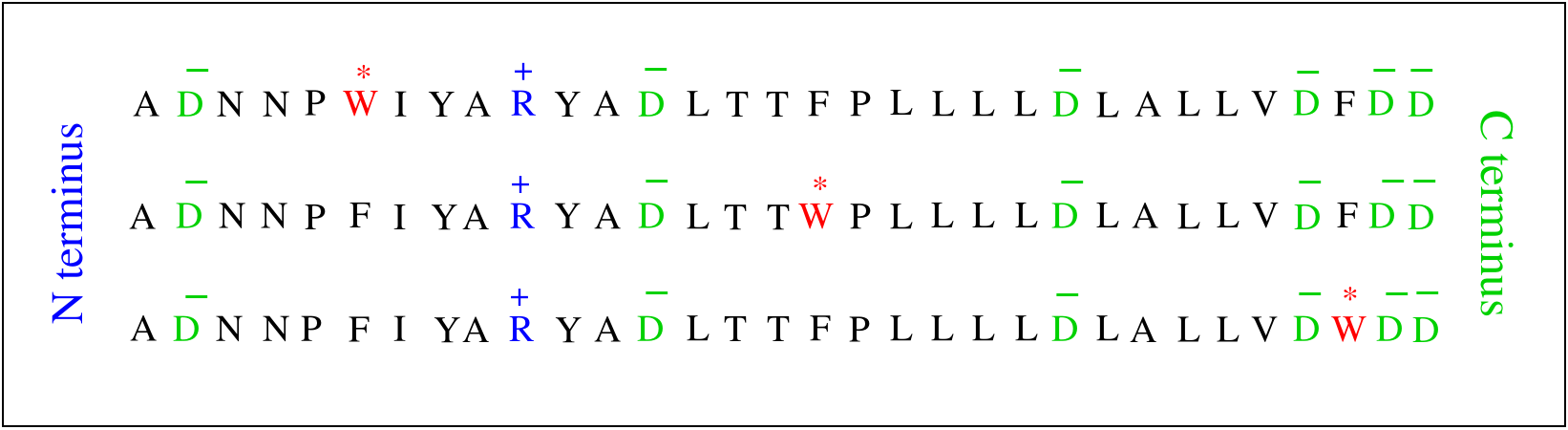}
\end{center}
\caption{Sequence of the $N_\mathrm{R}=32$ residues that make up variants W6, W17, W30 (top to bottom) of pHLIP. Highlighted are the residues \textsf{Arg}, \textsf{Asp}, which have electrically chargeable side chains, and the \textsf{Trp} residue used as a marker in fluorescent spectroscopy.}
  \label{fig:f9}
\end{figure}

The theoretical study reported here is custom-made for variants of the \emph{pH Low Insertion Peptide} (pHLIP) family \cite{RAS+08, AKW+10, KWW+12, WAR16, NWK+16, HSQ+16}, but not to the exclusion of other peptides with similar attributes.
The sequences of three pHLIP variants are shown in Fig.~\ref{fig:f9} with some relevant features highlighted.
The positioning in the sequence of charged residues and polar residues is instrumental for the solubility of pHLIP in water.
The hydrophobic residues provide the affinity for adsorption of pHLIP to the water-lipid interface.
The protonatable negative charges at and near the (inserting) \textsf{C} terminus make pHLIP sensitive to the experimentally controllable and reversible change in pH.

The \textsf{Trp} residues are markers for fluorescent spectroscopy, by which the insertion and exit processes are monitored \cite{KWW+12,TYM+09, RKB01}.
The three pHLIP variants with \textsf{Trp} at different positions in the sequence of residues are designed to yield clues about the modes of insertion and exit.
The conformational changes between coil and helix segments that accompany both insertion and exit can be monitored by circular dichroism (CD) spectroscopy \cite{KWW+12, VSS+18}.
The helix-inhibiting \textsf{Pro} residue near the center of the sequence may be instrumental for the mechanics of the insertion process, allowing a kink between two fully formed helical segments \cite{note2, Heij91, CBS02, WD91}. 
Peptide insertion into membranes is, of course, a wider field of research beyond the limited focus of this study.
There are aggregates of peptides which interact with membranes in significantly different ways, by forming pores, for example. 
In this work no peptide aggregates are being considered.

The following sections are about fields, profiles, landscapes, and pathways: fields of environmental parameters (Sec.~\ref{sec:mem-env}), profiles for local properties of long polypeptides (Sec.~\ref{sec:prof}), landscapes for global properties of short peptides (Sec.~\ref{sec:land}), and pathways associated with descending free energy of short peptides (Sec.~\ref{sec:inserpath}).
Finally, we briefly discuss effects not yet accounted for and outline the challenges facing kinetic studies (Sec.~\ref{sec:con-out}). 

%
\section{Membrane environment}\label{sec:mem-env}
%
All heterogeneity in a lipid bilayer considered here is associated with the normal spatial coordinate $x$.
We set $x=0$ at the center of the bilayer.
The outside of the cell or liposome is at positive $x$.
Any effects of curvature are set aside as higher-order corrections to results presented here.
The membrane environment is characterized by several parameters.
We take the dominant parameter field to be the concentration of $\mathrm{H_2O}$ molecules.
Hydrophobic interactions are prevalent \cite{TG12}.
Subdominant parameter fields involve electrostatic interactions including trans-membrane, surface, and dipole potentials \cite{Wang12}. 
Further parameters are related to properties of lipids, notably the profile of lateral pressure and the entropy reduction along the contact line with the peptide \cite{BenS95, Seel04, KIM08, LVA12, NWK+16}.
We examine the effects of the dominant environmental parameter in some detail and discuss those of subdominant parameters summarily at the end.

\subsection{Density field of water}\label{sec:den-fie-wat}
\begin{figure}[htb]
  \begin{center}
  \includegraphics[width=70mm]{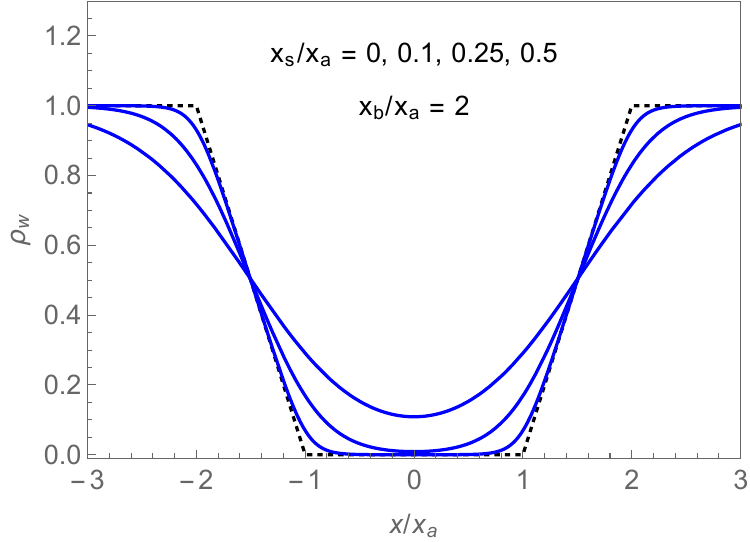}
\end{center}
\caption{Smoothed-ramp density field of $\mathrm{H_2O}$ molecules. The parameters $x_\mathrm{a}$ and $x_\mathrm{b}$ in (\ref{eq:1}) locate the bottom and the top of the ramp symmetrically, and $x_\mathrm{s}$ controls the softening of the edges.}
  \label{fig:f1}
\end{figure}

The dominant environmental parameter, the density field $\rho_\mathrm{w}(x)$ of $\mathrm{H_2O}$ molecules, is symmetric under reflection about $x=0$.
It is a dimensionless quantity varying between $\rho_\mathrm{w}(x)=1$ sufficiently far from the lipids and $\rho_\mathrm{w}(x)\ll1$ near the center of the bilayer.
We use a smoothed-ramp density field as a model representation in our statistical mechanical analysis:
\begin{align}\label{eq:1} 
\rho_\mathrm{w}(x) =
1 - \frac{x_{s}}{x_\mathrm{a}-x_\mathrm{b}}
      \ln\!\left(
        \frac{\cosh\left(\frac{x}{x_\mathrm{s}}\right)
        +\cosh\left(\frac{x_\mathrm{a}}{x_\mathrm{s}}\right)}
        {\cosh\left(\frac{x}{x_\mathrm{s}}\right)
        +\cosh\left(\frac{x_\mathrm{b}}{x_\mathrm{s}}\right)}\right)\!.    
\end{align}
It has two control parameters, $x_\mathrm{b}/x_\mathrm{a}>1$ and ${x_\mathrm{s}/x_\mathrm{a}>0}$.
A density field of such shape (Fig.~\ref{fig:f1}) is well-established from experiment \cite{ALE+03} and computer simulations \cite{MBT08}.

Throughout Secs.~\ref{sec:mem-env}-\ref{sec:land} we use the following specifications: $x_\mathrm{a}=15$\AA~represents the distance from the center of the bilayer to a point just inside the lipid headgroups, and $x_\mathrm{b}=25$\AA~the distance from the center to a point outside, where $\mathrm{H_2O}$ molecules are in contact with the polar ends of the lipid headgroups; the value $x_\mathrm{s}\simeq3$\AA ~ensures smoothness over an atomic length scale \cite{note3}.
The results presented in the following are not sensitive to small variations in these parameters.

A peptide of 32 residues in helix conformation would have a length of roughly $50$\AA, if we assume that each helical link \cite{note5} adds an element $l_\mathrm{h}\simeq1.5$\AA ~of length in the direction of the axis \cite{FP02}.
The length (end-to-end distance) of the peptide in the coil conformation is a fluctuating quantity.
With a size $l_\mathrm{a}\simeq4$\AA ~for each link, the contour length of the peptide becomes more than double its length in helix conformation.
The coil conformation, modeled as a random walk,
has an average end-to-end distance of {$\sim24$\AA} if unrestricted.
Geometrical and dynamical constraints make the average end-to-end distance considerably longer \cite{GK94}.

\subsection{Free-energy}\label{sec:free-ene}
The term \emph{free-energy landscape} of the title requires some explanation.
The system under consideration includes a peptide in an environment consisting of a lipid bilayer surrounded by water.
For a homogeneous system, the Gibbs free energy can be expressed in the form,
\begin{equation}\label{eq:36} 
G(T,P)=U+PV-TS=H-TS,
\end{equation}
where $U$ is the internal energy, $H$ the enthalpy, $S$ the entropy, and $V$ the volume. 
The pressure $P$ and the temperature $T$ are control variables.

Our system is not homogeneous in all respects.
We consider situations at uniform $T$, typically room temperature ($T_\mathrm{rm}=293$K).
The pressure is uniform in the aqueous environment and the normal pressure, $P_\mathrm{N}$, also across the bilayer.
However, the lateral pressure, $P_\mathrm{L}$, has a characteristic profile that averages out to the value of the normal pressure \cite{Mars96, Cant97, Cant99, GS04, GBM06, Mars07, BMT10, ZL13, DFST16, DM16}.
For the purpose of this study, we only consider quasistatic processes of a restricted type in which both $T$, and $P_\mathrm{N}$ remain constant and uniform.
One natural energy scale uses $k_\mathrm{B}T_\mathrm{rm}\simeq 4.0\times10^{-21}\mathrm{J} \simeq 0.58\mathrm{kcal/mol}$ as its unit.
The processes involve the translocation of the peptide, accompanied by changes in  conformation of the peptide and in its interactions with water and lipids.

Each quasistatic process can be described as a path in a space of independent variables.
The variation of $G$ ascends or descends in a landscape of sorts, just as a path on a topographic map does.
Path segments with $\Delta G<0$ are favorable and path segments with $\Delta G>0$ unfavorable regarding spontaeneous occurrence.
Paths that are all downhill are likely to have fast realizations in experiments.
By contrast, paths that have significant barriers between initial and final points are much slower if realized at all.

All changes of $G$ are a combination of enthalpic and entropic contributions.
For the processes under consideration here we can write,
\begin{equation}\label{eq:37} 
\Delta G=\Delta H - T\Delta S.
\end{equation}
We refer to changes $\Delta H<0$ as associated with an \emph{enthalpic gain} and changes $\Delta S>0$ as associated with an \emph{entropic gain} in the sense that a gain lowers $G$ while a loss does the opposite. 

Entropic losses, $\Delta S<0$, that are relevant  for this study include the following: 
(i) a complete or partial conformational change of the peptide from (disordered) coil to (ordered) helix;
(ii) the immobilization of $\mathrm{H_2O}$ molecules via the formation of H-bonds with polar contacts on the backbone or the side chains of the peptide or with polar contacts on lipid head-groups; 
(iii) the formation of an ordered contact line between lipid head-groups and the peptide in adsorbed or trans-membrane states.

Enthalpic losses related to a change $\Delta U>0$ are all associated with molecular interactions. 
The two main sources of this type in the context of this study involve (i) the breaking of H-bonds and (ii) the translocation of charges or polar contacts from a polar environment (water) into a nonpolar environment (membrane).
The H-bonds in question include internal ones between backbone contacts, and external H-bonds between the peptide (backbone or side chain), $\mathrm{H_2O}$ molecules, and lipid head-groups, all of which have polar contacts.
A different type of enthalpic loss, $(P\mathrm{_L^h}-P\mathrm{_L^l})\Delta V>0$, comes into play when a peptide segment (e.g. a residue) of volume $\Delta V$ translocates from a position of low lateral pressure, $P\mathrm{_L^l}$, to a position of high lateral pressure, $P\mathrm{_L^h}$.

It is quite challenging, in general, to estimate all these contributions with some accuracy. 
Existing estimates found in the literature vary widely, in part due to differences in underlying assumptions.
In what follows, a case will be made for identifying and quantifying dominant contributions.

\subsection{Enthalpic cost of H-bonds}\label{sec:enth-cost}
In the $\alpha$-helix conformation the backbone of each residue participates in two H-bonds.
The CO group of residue $n$ is acceptor to the NH group of residue $n+4$ acting as donor.
The helix conformation thus involves one internal H-bond per residue.
The conversion of a helix segment into a coil segment breaks internal H-bonds, for which there is an enthalpic cost.

Deep inside the lipids the full price is due, up to 5kcal/mol per H-bond, which is considerable in units of ambient thermal fluctuations.
When the peptide is positioned in the polar environment of the lipid headgroups and the adjacent water, there are opportunities for internal H-bonds to be replaced by external ones.
The replacement bonds reduce the (enthalpic) cost of breaking internal H-bonds.
At the same time, it increases the maximum number of H-bonds per residue from one to two.
The replacement bonds (with reduced directionality) are likely to be weaker.

Whether the enthalpic bottom line in this case is a gain or loss depends on how the exposed backbone of the polypeptide affects the internal H-bonds of liquid water.
In ice there are two intact H-bonds per $\mathrm{H_2O}$ molecule.
Each molecule shares four bonds, two in the role of donor and two as acceptor.
In liquid water about 12\% of H-bonds between $\mathrm{H_2O}$ molecules are broken \cite{FP02, note6}.
The intact H-bonds form a dynamic network with $\mathrm{H_2O}$ molecules, sharing less than four bonds on average.

If the fraction of unsatisfied H-bonds between $\mathrm{H_2O}$ molecules is higher at the lipid-water interface than inside the bulk, then the exposed CO groups and NH groups along a coil segment of the backbone are more likely to encounter partners for external H-bonds in adjacent $\mathrm{H_2O}$ molecules.
This reduces the need for breaking H-bonds between $\mathrm{H_2O}$ molecules when external H-bonds are formed along the backbone of the peptide.
The formation of external H-bonds in water is then more likely to result in an enthalpic gain.

The methodology developed in Ref.~\cite{cohetra} encodes the enthalpic contribution to the conformational affinity in the activation energy $\epsilon$ of a coil link from the helix reference state. 
The heterogeneous membrane environment is accounted for by turning this parameter into a field.
We use the ansatz,
\begin{equation}\label{eq:2} 
\epsilon(x)=\epsilon_\mathrm{Hb}\big[1-\alpha_\mathrm{H}\rho_\mathrm{w}(x)\big],
\end{equation}
for the dependence on position, the premise being that the density field $\rho_\mathrm{w}(x)$ of water represents the dominant environmental influence.
Near the center of the lipid bilayer we have $\rho_\mathrm{w}(x)\ll1$, which maximizes $\epsilon(x)$ to roughly the strength of an internal H-bond.
At positions closer to the lipid-water interface, $\epsilon(x)$ decreases as $\rho_\mathrm{w}(x)$ increases.
This change is due to the growing probability that internal H-bonds are being replaced by external ones.

The parameter $\alpha_\mathrm{H}$ determines whether in the aqueous environment, with $\rho_\mathrm{w}(x)\simeq1$, we have an enthalpic gain $(\alpha_\mathrm{H}>1)$ or an enthalpic loss $(\alpha_\mathrm{H}<1)$.
Here we use $\epsilon_\mathrm{Hb}/k_\mathrm{B}T\simeq9$, i.e. $\epsilon_\mathrm{Hb}\simeq5$kcal/mol. The strength of H-bonds varies and depends on geometrical constraints and the charge distribution on the polar contacts. H-bonds between $\mathrm{H_2O}$ molecules in liquid water are highly fluctuating in strength, and could be as low as 2kcal/mol on average \cite{SYSS03, FR05, LH06, JMF+08, GBPK09, CHSB17}. 

\subsection{Entropic cost of H-bonds}\label{sec:entr-cost}
The enthalpic cost reduction associated with external H-bonds in the polar environment of liquid water comes with an entropic price that has yet to be included in the accounting. 
Every H-bond formed between an exposed backbone CO or NH group with an $\mathrm{H_2O}$ molecule reduces the mobility of that water molecule to some extent and thus lowers its contribution to the entropy.

It is hard to estimate the magnitude $|\Delta \bar{S}_\mathrm{H}|/k_\mathrm{B}$ of this entropy reduction from first principles. 
At a glance, we might expect it to be comparable in magnitude to the entropic gain per residue produced when a segment of (ordered) helix transforms into a segment of (disordered) coil.
In Ref.~\cite{cohetra} we have calculated that entropy gain per residue to range between $ \ln2\simeq0.7$ and $\ln3\simeq1.1$ in units of $k_\mathrm{B}$.
Backbone segments in coil conformation are less mobile than $\mathrm{H_2O}$ molecules in the dynamic network of H-bonds that make up liquid water.
The unstructured nature of the coil reduces the entropy loss of $\mathrm{H_2O}$ molecules forming H-bonds with it.
All this makes it reasonable to operate with an estimated upper limit,
\begin{equation}\label{eq:3} 
\frac{|\Delta \bar{S}_\mathrm{H}|}{k_\mathrm{B}}\lesssim 1.
\end{equation}
The actual value of $|\Delta \bar{S}_\mathrm{H}|$ could be lower on account of the role played by the lipid headgroups.
The zwitterionic headgroups of POPC, for example, offer alternative contacts for external H-bonds to backbone segments in coil conformation. 
They come at lower entropic cost.

The overall message for what follows in Sec.~\ref{sec:land} is that the entropic contribution to the free energy of the peptide backbone in coil conformation is significantly reduced from what the disorder of the coil conformation would suggest when the peptide is in contact with water.

\subsection{Model for peptide conformation}\label{sec:mod-pep-conf}
In Ref.~\cite{cohetra}, we exactly solved a microscopic statistical mechanical model for the conformational transformation between coil and helix of a long polypeptide positioned in a plane parallel to the interface of a polar and a nonpolar medium such as realized by water and a lipid bilayer.
Depending on the settings and variations of its parameters, the model predicts a conformational crossover or a transition of first or second order between coil and helix.

\begin{figure}[htb]
  \begin{center}
\includegraphics[width=84mm]{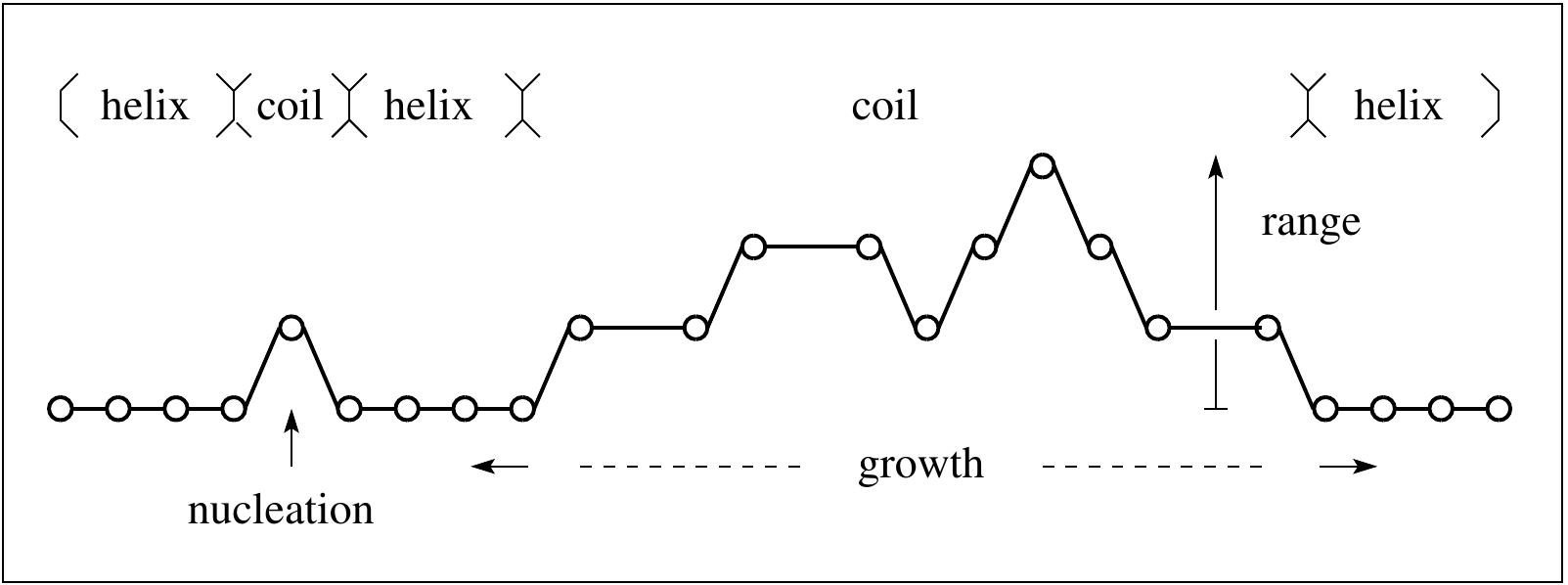}
\end{center}
\caption{Coil conformation of polypeptide modeled as a self-avoiding random walk generated by statistically interacting nested particles activated from the helix reference state via \emph{nucleation} (control parameter $\tau$) and \emph{growth} (control parameter $t$) with \emph{range} limited by control parameter $\mu$. }
  \label{fig:f11}
\end{figure}

All microstates of the peptide are characterized, in this model, by combinations of $2\mu$ species of statistically interacting nested particles: hosts $(m=1)$ nucleate coil segments, whereas hybrids $(m=2,\ldots,\mu)$ and tags $(m={\mu+1},\ldots,2\mu)$ grow such segments in two perpendicular directions of a self-avoiding random walk as illustrated in Fig.~\ref{fig:f11}.

The activation energies $\epsilon_m$, $m=1,\ldots,2\mu$, can be tailored to meet the physical requirements at hand.
In the plane of the water-lipid interface all activation energies are spatially uniform.
Nucleating a coil segment requires that several internal H-bonds along the backbone of the $\alpha$-helix are broken simultaneously, whereas the growth of a coil segment already nucleated proceeds by the sequential breaking of single H-bonds.

The model, therefore, assigns different activation energies for the control of nucleation and for the control of growth, namely $\epsilon_1\doteq\epsilon_\mathrm{n}$ for hosts, $\epsilon_2=\cdots=\epsilon_\mu \doteq 2\epsilon_\mathrm{g}$ for hybrids, and $\epsilon_{\mu+1}=\cdots=\epsilon_{2\mu}\doteq\epsilon_\mathrm{g}$ for tags.
These activation energies in units of the thermal energy $k_\mathrm{B}T$ are usefully expressed by the nucleation parameter $\tau$ (also named cooperativity) and growth parameter $t$:
\begin{subequations}\label{eq:4}
\begin{align}\label{eq:4a} 
& \tau=e^{\beta(\epsilon_\mathrm{g}-\epsilon_\mathrm{n})}\quad :~ 0<\tau\leq1, \\
& t=e^{\beta\epsilon_\mathrm{g}}\quad \hspace{7.5mm}:~ 0< t<\infty.
\end{align}
\end{subequations}
In addition to these two continuous control parameters, the discrete model parameter $\mu$ controls the range of the random walk away from the axis of the local helix segments.
All model features from Ref.~\cite{cohetra} used in the following are reviewed in Appendix~\ref{sec:appa}.

\subsection{Model parameter field}\label{sec:par-fie-pep}
Of the three model parameters, the growth parameter $t$ is the one most sensitive to the environment by far.
We begin the model adaptation to the membrane environment by keeping the cooperativity parameter $\tau$ and the range parameter $\mu$ uniform, while we turn $t$ into a field.
For this purpose, we use the ansatz (\ref{eq:2}) as we construct two fields of scaled activation energy, $\epsilon_\mathrm{n}(x)$ for hosts, $\epsilon_\mathrm{g}(x)$ for hybrids and tags \cite{note8}.
We leave the former $(m=1)$ undetermined for now and link the latter $(m=2,3,\ldots,2\mu)$ to the density field of water:
\begin{align}\label{eq:5} 
& K_\mathrm{n}(x)\doteq\frac{\epsilon_\mathrm{n}(x)}{k_\mathrm{B}T},
\\ \label{eq:6} 
& K_\mathrm{t}(x)\doteq \frac{\epsilon_\mathrm{g}(x)}{k_\mathrm{B}T}=
 \frac{\epsilon_\mathrm{Hb}}{k_\mathrm{B}T}\big[1-\alpha_\mathrm{H}\rho_\mathrm{w}(x)\big], 
\end{align}
where $\epsilon_\mathrm{Hb}/k_\mathrm{B}T$ represents the scaled energy of an H-bond and $\alpha_\mathrm{H}\simeq1$ is the enthalpy parameter introduced previously (Sec.~\ref{sec:enth-cost}), assumed to be equal for hybrids and tags.
The growth parameter field,
\begin{equation}\label{eq:7} 
t(x)=e^{K_\mathrm{t}(x)}\quad :~0< t(x)<\infty,
\end{equation}
is environmentally sensitive via the shape of $\rho_\mathrm{w}(x)$ and the value of $\alpha_\mathrm{H}$.
The cooperativity,
\begin{equation}\label{eq:8} 
\tau=e^{K_\mathrm{t}(x)-K_\mathrm{n}(x)}
=t(x)e^{-K_\mathrm{n}(x)}\quad :~ 0<\tau\leq1,
\end{equation}
is kept as a position-dependent parameter.
The function $K_\mathrm{n}(x)$ is determined from (\ref{eq:8}).

This choice of modeling is informed by the following facts. 
Cooperativity, which controls the nucleation of coil segments, is a process initiated by thermal fluctuations within the backbone of an intact segment of $\alpha$-helix. 
Multiple H-bonds must be broken simultaneously.
They are all protected from direct environmental contact. 
Nucleation is only affected indirectly by an environmental change from nonpolar to polar.
In the nonpolar environment, the nucleation energy barrier is followed by a high plateau and in the polar environment by a low plateau.
The former favors a reversal of nucleation events, whereas the latter favors the growth of nucleated segments.

\begin{figure}[htb]
  \begin{center}
\includegraphics[width=40mm]{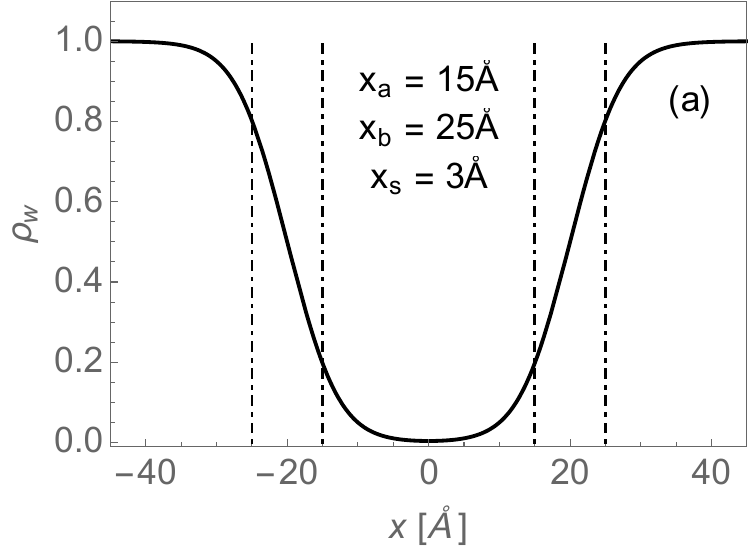}\hspace*{3mm}\includegraphics[width=40mm]{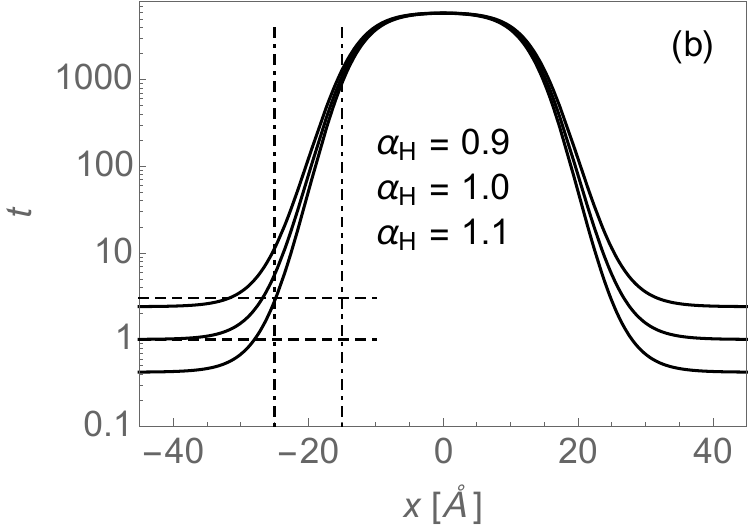}
\end{center}
\caption{(a) Model density field of water $\rho_\mathrm{w}(x)$ and (b) growth parameter field $t(x)$, with specifications as indicated. The dot-dashed lines represent the endpoints of the ramp and the dashed lines the range $1\leq t\leq3$ in which a coil-helix phase transition may occur.}
  \label{fig:f3}
\end{figure}

Figure~\ref{fig:f3}(a) shows the model density field of water $\rho_\mathrm{w}(x)$ used henceforth. 
It has the smoothed-ramp profile (\ref{eq:1}).
The growth parameter field $t(x)$ inferred from the predominant environmental field $\rho_\mathrm{w}(x)$ via (\ref{eq:6}) and (\ref{eq:7}) is shown in Fig.~\ref{fig:f3}(b).
When the polypeptide is in coil conformation while adsorbed to the water-lipid interface, its position is near the outer dot-dashed line.
For the coil conformation to be stable, the growth parameter $t(x)$ must be below the upper horizontal dashed line (at $t=3$).
That is barely the case if $\alpha_\mathrm{H}<1$.
A robust coil conformation requires that $\alpha_\mathrm{H}\gtrsim1$, implying that  breaking internal H-bonds along the backbone of the polypeptide and replacing them by external H-bonds with available $\mathrm{H_2O}$ molecules is an enthalpic gain.

The extended model system is now ready for applications to the heterogeneous membrane environment. 
We have converted $t$ into the field (\ref{eq:7}) and kept the control parameters $\tau$ and $\mu$ uniform.
We have already stated reasons for not turning the nucleation parameter $\tau$ into a field. 
Regarding the discrete parameter $\mu$ \cite{note9}, we will consider the cases $\mu=2$ and $\mu=\infty$, for which analytic solutions are available in Ref.~\cite{cohetra}. 
The two values span a range that is correlated with a range of entropy generated inside coil segments of given length.

%
\section{Profiles}\label{sec:prof}
%
The analysis reported in the following yields profiles for specific position-dependent attributes of a polypeptide backbone in the heterogeneous membrane environment.
The relevant quantities are the helicity $\bar{N}_\mathrm{hl}$ as well as densities of free energy $\bar{G}$, enthalpy $\bar{H}$, and entropy $\bar{S}$.
What is taken into account at this stage are the internal H-bonds along the backbone of a long, generic polypeptide and external H-bonds with water or lipid headgroups depending on their local availability.
Also taken into account is the entropy of the backbone in coil conformation and an entropic contribution associated with external H-bonds (Sec.~\ref{sec:entr-cost}).

\subsection{Internal H-bonds}\label{sec:iHb}
\begin{figure}[b]
  \begin{center}
\includegraphics[width=40mm]{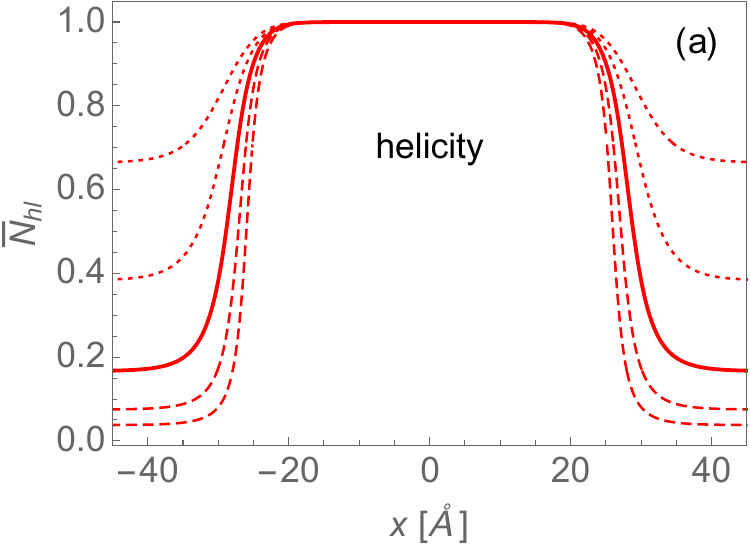}\hspace*{3mm}\includegraphics[width=40mm]{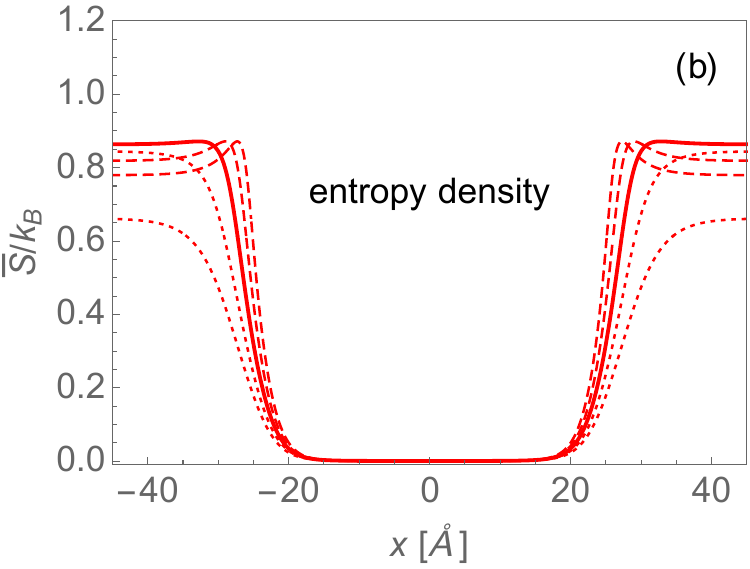}
\includegraphics[width=40mm]{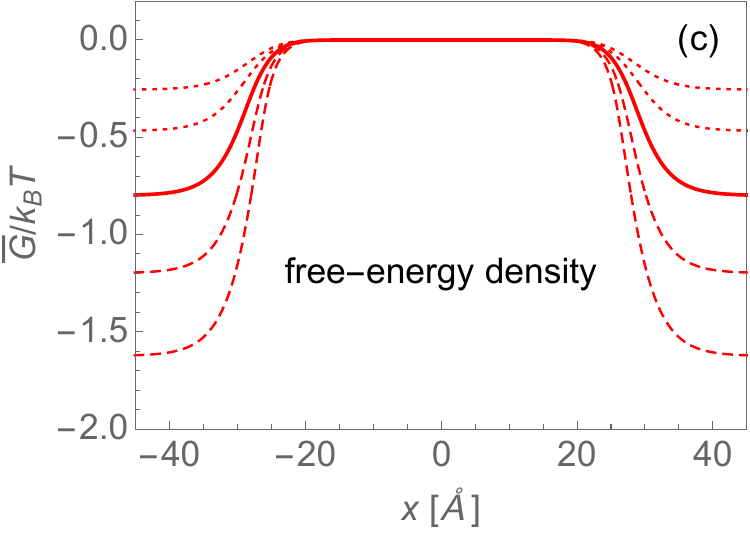}\hspace*{3mm}\includegraphics[width=40mm]{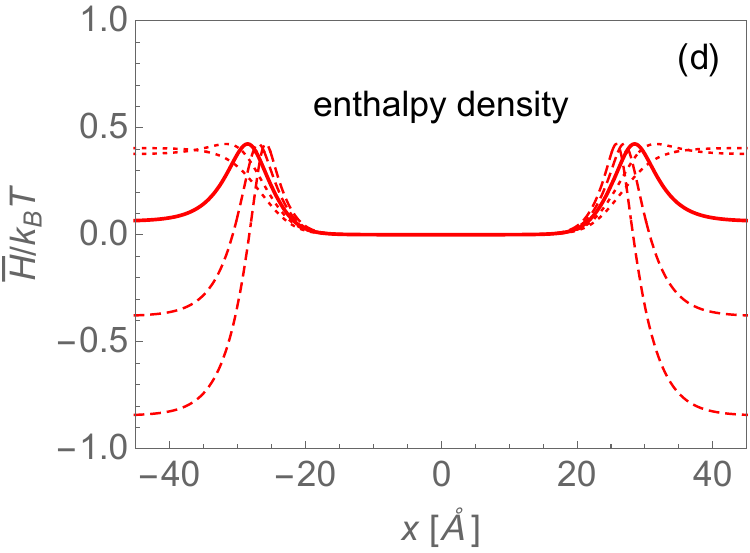}
\end{center}
\caption{Profiles of (a) helicity, (b) entropy density, (c) free-energy density, and (d) enthalpy density for a long polypeptide. The model parameter values are $\mu=2$ and $\tau=0.5$. The growth parameter field $t(x)$ uses (\ref{eq:6})-(\ref{eq:7}) with $\rho_\mathrm{w}(x)$ from Fig.~\ref{fig:f3}.
The solid, dashed, and dotted curves pertain to $\alpha_\mathrm{H}=1$, $\alpha_\mathrm{H}=1.05, 1.1$, and $\alpha_\mathrm{H}=0.95,0.9$, respectively.}
  \label{fig:f5}
\end{figure}

\begin{figure}[t]
  \begin{center}
\includegraphics[width=40mm]{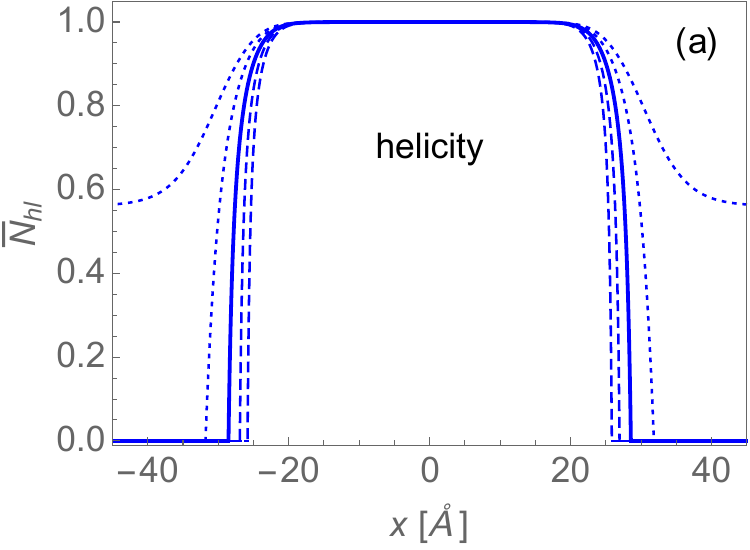}\hspace*{3mm}\includegraphics[width=40mm]{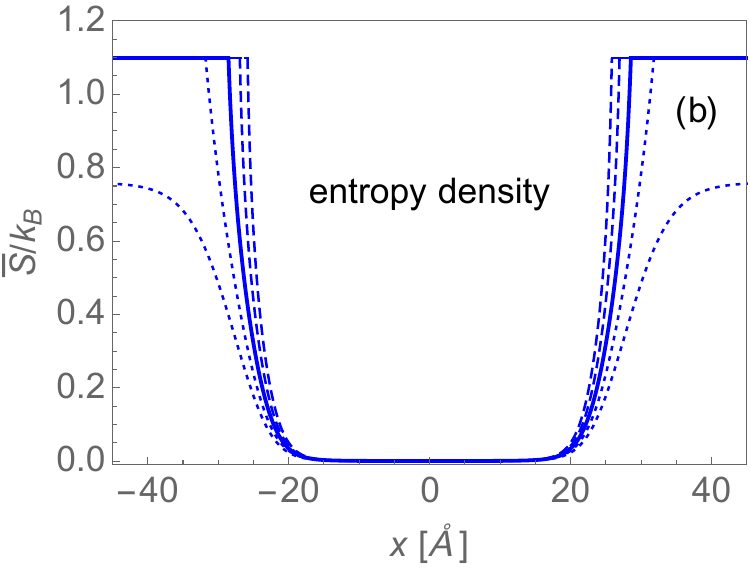}
\includegraphics[width=40mm]{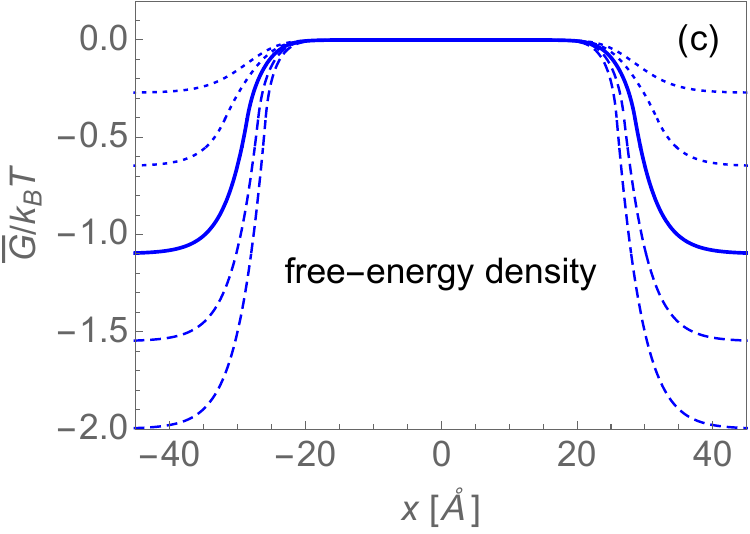}\hspace*{3mm}\includegraphics[width=40mm]{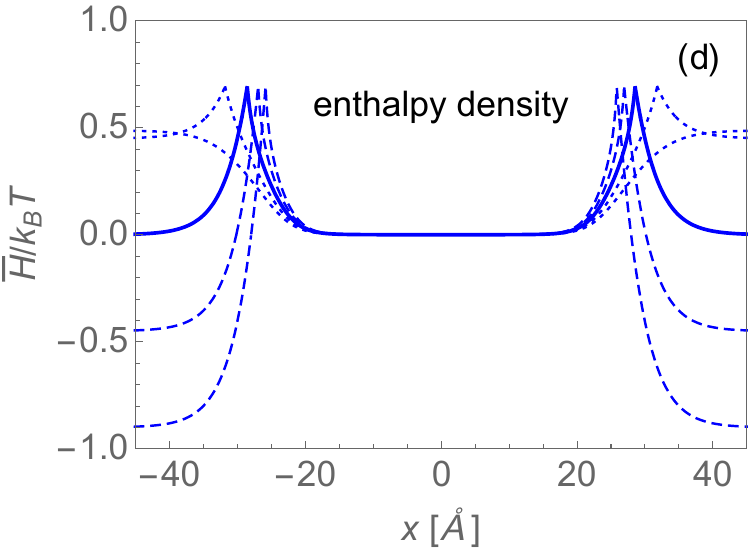}
\end{center}
\caption{Profiles of helicity, entropy density, free-energy density, and enthalpy density for a long polypeptide. The model parameter values are $\mu=\infty$ and $\tau=0.5$. The growth parameter field $t(x)$ uses (\ref{eq:6})-(\ref{eq:7}) with $\rho_\mathrm{w}(x)$ from Fig.~\ref{fig:f3}.
The solid, dashed, and dotted curves pertain to $\alpha_\mathrm{H}=1$, $\alpha_\mathrm{H}=1.05, 1.1$, and $\alpha_\mathrm{H}=0.95,0.9$, respectively.}
  \label{fig:f6}
\end{figure}

We construct profiles from the expressions for helicity $\bar{N}_\mathrm{hl}(t,\tau)$, entropy density $\bar{S}(t,\tau)$, free-energy density $\bar{G}(t,\tau)$, and enthalpy density $\bar{H}(t,\tau)$ stated in Appendix~\ref{sec:appa}, in combination with the field $t=t(x)$ for the growth parameter (\ref{eq:7}), and a value of choice for the nucleation parameter $\tau$, Eq.~(\ref{eq:8}). 
The growth parameter field $t(x)$ determines all profiles via local relations. 
Profiles for $\mu=2$ (narrow range of conformational disorder) are shown Fig.~\ref{fig:f5} and profiles for $\mu=\infty$ (broad range) in Fig.~\ref{fig:f6}.

Well inside the lipid bilayer, at $x\simeq0$, the helix conformation is firmly established.
All internal H-bonds are intact.
There is no configurational disorder.
Therefore, the order parameter (helicity) is close to saturation whereas the densities of enthalpy \cite{note10} and entropy are near zero.
In consequence, the free-energy density of the polypeptide rises only imperceptibly above its (zero) reference value as well.
At positions away from the center of the bilayer, the helicity decreases and the entropy density increases, the former reflecting a drop in (helical) order and the latter a rise in (coil-like) disorder, both associated with the same conformational change.

These conclusions are not without caveat.
Charged residues (e.g. \textsf{Arg}) tend to drag water into the membrane.
Aggregates of peptides can form a polar environment inside the membrane in different ways, starting with water bridges and ending in pores, for example \cite{DA07, Alle07, LVA13}

In Ref.~\cite{cohetra} we identified one source of order and two sources of disorder involving the secondary structure of the polypeptide backbone alone.
Order increases with the growth of segments of helix conformation.
Disorder is contained (i) in the spatial distribution of boundaries between segments of coil and helix and (ii) inside each (unstructured) coil segment.

The enthalpy density produces an energy barrier at locations near the lipid-water interface [Figs.~\ref{fig:f5}(d) and \ref{fig:f6}(d)]. 
Near the peak position the thermal fluctuations are just strong enough to break internal H-bonds, but the environment is not yet sufficiently polar to produce an adequate supply of external H-bonds as replacements.
The enthalpy density decreases (for different reasons) on each side of the peak.
On one side, broken bonds are rare, on the other side, they are energetically inexpensive.
Outside the interface, the enthalpy profile levels off in a high or low plateau depending on the value of the physical parameter $\alpha_\mathrm{H}$. 
That parameter also affects the drop in helicity and the rise in entropy.
Enthalpic loss $(\alpha_\mathrm{H}<1)$ favors order and suppresses disorder (for the peptide) in the aqueous environment.

There are some qualitiative and some quantitative differences between the curves for $\mu=2$ and for $\mu=\infty$.
The case $\mu=\infty$ produces pure coil with maximum entropy in the aqueous environment.
Pure coil conformation means zero helicity.
Coil segments generate significantly more entropy for $\mu=\infty$ than for $\mu=2$.
The enthalpic barriers near the interface are more pronounced in the case $\mu=\infty$.
This difference is attributable to an entropic effect.
The breaking of an H-bond at significant enthalpic cost is more likely to happen if the entropy produced is large $(\mu=\infty)$ than if it is small $(\mu=2)$.

The free-energy profiles in Figs.~\ref{fig:f5} and \ref{fig:f6} tell us that the incentives for the insertion of peptides must come from a source other than what has already been taken into account. 
The free-energy density $\bar{G}(x)$ is significantly higher in the membrane environment than in the surrounding water. 
The dependences of $\bar{H}(x)$ and $\bar{S}(x)$ on $\alpha_\mathrm{H}$ are strong, but not decisively so.
The entropic contributions accounted for thus far are dominating the free energy.
Coil is favored over helix. Water beats lipids as the favored environment for the peptide.

\subsection{External H-bonds}\label{sec:eHb}
\begin{figure}[htb]
  \begin{center}
\includegraphics[width=40mm]{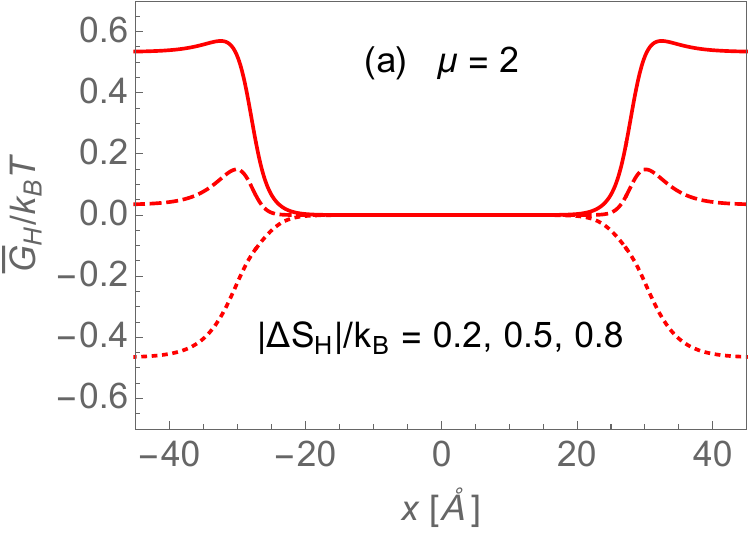}\hspace*{3mm}%
\includegraphics[width=40mm]{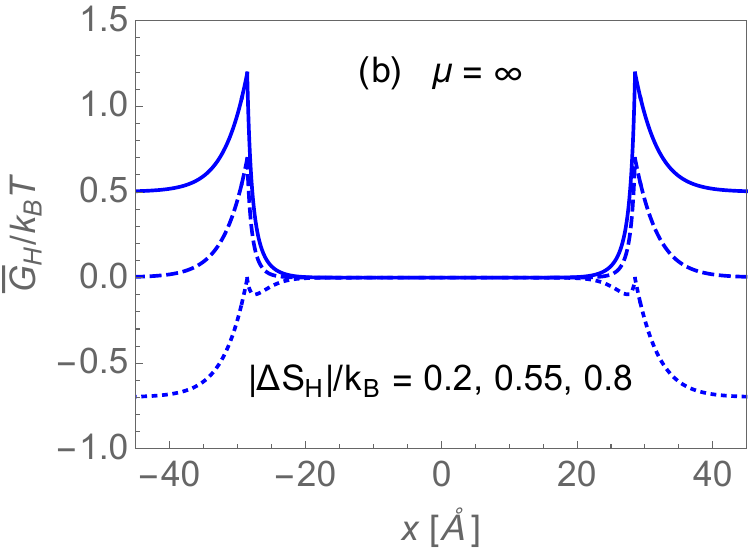}
\end{center}
\caption{Profiles of free-energy density (\ref{eq:9}) which includes an entropic contribution from external H-bonds. The two panels pertain to different values of the range parameter $\mu$. The curves from bottom to top are for increasing values of $|\Delta\bar{S}_\mathrm{H}|/k_\mathrm{B}$.  Additional model parameter values are $\tau=0.5$ and $\alpha_\mathrm{H}=1$.  Note the different vertical scales left and right.}
  \label{fig:f7}
\end{figure}

At this point in the analysis, we add one other contribution to the free-energy density profile.
Further contributions, associated with side chains and their interaction with lipids, will be introduced in Sec.~\ref{sec:side-chain}.
The contribution discussed here is entropic in nature and favors insertion.
The replacement of backbone internal H-bonds with external H-bonds that immobilize $\mathrm{H_2O}$ molecules from the aqueous environment, while providing an enthalpic discount as already accounted for, comes at an entropic cost \cite{DMA+10}.
This effect is taken into account via an amended free-energy density constructed as follows:
\begin{equation}\label{eq:9} 
\frac{\bar{G}_\mathrm{H}(x)}{k_\mathrm{B}T}=\frac{\bar{G}(x)}{k_\mathrm{B}T}+2\frac{|\Delta \bar{S}_\mathrm{H}|}{k_\mathrm{B}}\Big[1-\bar{N}_\mathrm{hl}(x)\Big],
\end{equation}
with $|\Delta S_\mathrm{H}|$ as discussed and estimated in Sec.~\ref{sec:entr-cost}.
The factor in square brackets represents the fraction of backbone segments in coil conformation with each segment offering docks for two $\mathrm{H_2O}$ molecules.
Profiles for the amended free-energy densities are shown in Fig.~\ref{fig:f7}.

This amendment does indeed contribute an incentive for insertion.
Its impact depends on the size of $|\Delta \bar{S}_\mathrm{H}|/k_\mathrm{B}$, which controls the entropy reduction due to external H-bonds, and on the range parameter $\mu$, which controls entropy production of coil segments.
An increase of $|\Delta \bar{S}_\mathrm{H}|/k_\mathrm{B}$ switches the global minimum in the free-energy density from the exterior region to the interior region. 
The switch happens at $|\Delta \bar{S}_\mathrm{H}|/k_\mathrm{B}\simeq0.5$ for $\mu=2$. 
A somewhat larger value, $|\Delta \bar{S}_\mathrm{H}|/k_\mathrm{B}\simeq0.55$ is needed to cause the switch for $\mu=\infty$.
These values are well below the estimated upper bound (\ref{eq:3}).
Unsurprisingly, the larger range parameter, which produces more backbone entropy, requires more entropy reduction in compensation before it yields an insertion incentive.

Interestingly, an energy barrier between the exterior and interior levels of free-energy-density builds up as $|\Delta \bar{S}_\mathrm{H}|/k_\mathrm{B}$ increases in size. 
This barrier, which is more pronounced in panel (b), helps stabilize the (experimentally confirmed) coexistence of short peptides in states of solution and adsorption.
Energy barriers such as emerge here quite naturally, play an important role in kinetic studies of the insertion process.

%
\section{Landscapes}\label{sec:land}
%
In the continuation of the analysis, we interpret the profiles calculated in Sec.~\ref{sec:prof} as propensities of backbone segments of short peptides such as pHLIP. 
The helicity profile and the density profiles for enthalpy, entropy, and free energy are employed here as one factor affecting the behavior of residues of short peptides in the same environment, specifically their conformational preference (coil or helix).
Other factors depend on attributes of the specific side chains and on further (enthalpic and entropic) effects of peptide-lipid interactions.

Landscapes of interest include those of free energy, enthalpy, entropy, and helicity.
It would take a molecular dynamics simulation to explore entire landscapes, i.e. the full range of configurations that a peptide can assume in the membrane environment.
The approach taken here, which is more limited in scope (see Sec.~\ref{sec:inserpath}), is more selective with configurations.
This selectivity is justifiable, at least in part, by the fact that the range of configurations is naturally and severely restricted by conformational constraints, specifically by the rigidity of the helix conformation.

\subsection{Three-variable landscapes}\label{sec:scen-l2}
\begin{figure}[htb]
  \begin{center}
  \includegraphics[width=75mm]{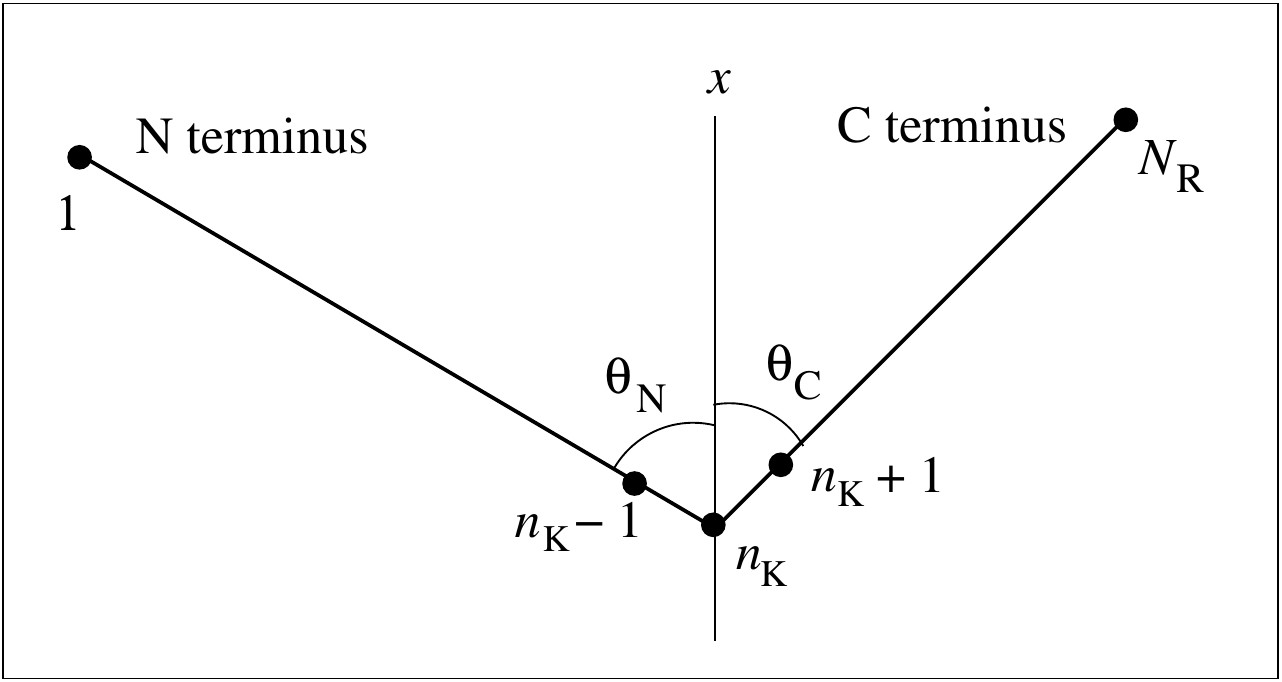}
\end{center}
\caption{Variables  $x_{n_\mathrm{K}}$ (depth of kink) and $\theta_\mathrm{N}$, $\theta_\mathrm{C}$ (angles between normal and segments). The counting of residues begins at the \textsf{N} terminus as is custom.}
  \label{fig:f12}
\end{figure}

In the following, we investigate three-variable free-energy landscapes for short peptides in varying positions and orientations.
The model peptide has $N_\mathrm{R}$ residues and is assumed to consist of two straight segments as schematically represented in Fig.~\ref{fig:f12}.
In applications to pHLIP, a likely candidate for the kink position is the helix inhibiting \textsf{Pro} residue \cite{Heij91, CBS02, WD91}.
One simulation study \cite{VSS+18} places a kink at the position of the \textsf{Asp} residue which is somewhat closer to the \textsf{N} terminus than the \textsf{Pro} residue.

We use the position $x_{n_\mathrm{K}}$ of the kink on the normal to the bilayer and the angles $\theta_\mathrm{N}$, $\theta_\mathrm{C}$ of the segments ending in the \textsf{N}, \textsf{C} terminus, respectively, as the variables that specify the position and orientation of the peptide in the membrane environment.
The lengths of both segments depend on the local conformation, which, in turn, depend on the location of a given segment in the membrane environment as specified below.

We intend to calculate the free energy of peptides along specific pathways in this three-variable landscape. 
Our focus will be on the pHLIP variants with the sequences as stated in Fig.~\ref{fig:f9}.
We consider three principal states I, II, and III of pHLIP in the membrane environment and specific pathways between them \cite{RSA+07, RMAE20}:

I: pHLIP is in aqueous solution and in coil conformation. 
We shall start adsorption pathways at coordinates $x_{n_\mathrm{K}}=27$\AA, $\theta_\mathrm{N}=\theta_\mathrm{C}=90^\circ$. 

II: pHLIP is adsorbed to the outside interface of a cell membrane with interstitial fluid or of a liposome with water. 
At high pH the adsorption is rather loose and the conformation is coil.
At low pH the adsorption is deeper and the conformation is largely $\alpha$-helix. 
We  shall see that the adsorbed state has the kink of the peptide positioned deeper in the membrane and both termini sticking out into the water: $x_{n_\mathrm{K}}\simeq17$\AA, $\theta_\mathrm{N}\simeq \theta_\mathrm{C}\simeq30^\circ$.

III: pHLIP is in a trans-membrane state with a helical central part and short coil-like flanking parts. The coordinates roughly are $x_{n_\mathrm{K}}\simeq0$, $\theta_\mathrm{N}\simeq0$, ${\theta_\mathrm{C}\simeq180^\circ}$.

In order to calculate the peptide free energy in any of the three principal states and in states that connect them we need to know the position $x_n$ of every residue in the membrane environment.
We determine these positions recursively from distances between residues beginning at the kink.
Writing,
\begin{align}\label{eq:29}
& x_{n-1}=x_n+l(x_n)\cos\theta_\mathrm{N} \quad 
:~ n=n_\mathrm{K},\ldots,2,\\ \nonumber
& x_{n+1}=x_n+l(x_n)\cos\theta_\mathrm{C} \quad
:~ n=n_\mathrm{K},\ldots,N_\mathrm{R}-1,
\end{align}
takes into account that the distance between adjacent residues depends on the local conformation of the residue position in the membrane environment.
That conformation is either coil or helix with probabilities for which we use the propensity profile as calculated in Sec.~\ref{sec:prof}:
\begin{align}\label{eq:16} 
x_{n+1}-x_n\doteq l(x_n)
= l_\mathrm{c}-(l_\mathrm{c}-l_\mathrm{h})\bar{N}_\mathrm{hl}(x_n).
\end{align}
Whereas the length of a helical segment is given, $l_\mathrm{h}=1.5$\AA, the  (averaged) length $l_\mathrm{c}$ of coil segments becomes a physical model parameter.
Its maximum value is the contour length per residue of the backbone: $l_\mathrm{c}\lesssim 4$\AA.

\subsection{Backbone contributions}\label{sec:bb-con}
It is useful to look at some free-energy and helicity landscapes that represent the effects of the backbone alone.
This is facilitated in Appendix~\ref{sec:appc}.
The main message from the results presented there to what follows is twofold: (i) The backbone-lipid interaction favors coil conformation near the water-lipid interface and helix conformation inside the membrane. (ii) Insertion lowers the backbone contribution to the free energy. 

Here we merely state the expression for the backbone contribution to the free-energy landscape and the expression for the helicity (fraction of peptide in helix conformation) to be used in the following:
\begin{align}\label{eq:30}
& G_\mathrm{BB}(x_{n_\mathrm{K}},\theta_\mathrm{N},\theta_\mathrm{C})=
\sum_{n=1}^{N_\mathrm{R}}
\bar{G}_\mathrm{H}(x_n),
\\ \label{eq:31}
& N_\mathrm{hl}^{(\mathrm{K})}(x_{n_\mathrm{K}},\theta_\mathrm{N},\theta_\mathrm{C})=\sum_{n=1}^{N_\mathrm{R}}\bar{N}_\mathrm{hl}(x_n),
\end{align}
where the function $\bar{G}_\mathrm{H}(x)$ is taken from (\ref{eq:9}) and the functions $\bar{G}(x)$, $\bar{N}_\mathrm{hl}(x)$ from Appendix~\ref{sec:appa}.
The dependence of $x_n$ on $x_{n_\mathrm{K}},\theta_\mathrm{N},\theta_\mathrm{C}$ is given in (\ref{eq:29}).

\subsection{Side-chain contributions}\label{sec:side-chain}
The contributions to the free-energy landscape originating from the side chains of a peptide with a given sequence of residues are manifold \cite{NLS66}.
Their interactions with lipid molecules include aspects of hydrophobicity, pressure differentials, entropy reductions, and electrostatics.
Ranking the relative importance has been challenging and not without controversy.
For the adsorption and insertion pathways presented in Sec.~\ref{sec:inserpath} we use  well-established transfer-free-energy data as the dominant side-chain contribution.
The potential impact of other contributions will be discussed in Sec~\ref{sec:con-out} as an outline of future work.

The side chains of pHLIP residues range from strongly hydrophobic to strongly hydrophilic. 
Their transfer between polar and nonpolar environments contributes significantly to the free-energy landscape.
Some side chains carry (positive or negative) electric charges.
The protonation at low pH of the negatively charged \textsf{Asp} residues and \textsf{C} terminus (see Fig.~\ref{fig:f9}) changes the overall hydrophobicity critically as we shall see, destabilizing the adsorbed state II in favor of the trans-membrane state III.
An estimate of the side-chain transfer-free energy between states I, II, and III based on the Wimley-White interface and octanol scales is given in Appendix~\ref{sec:appd}. 
That scheme takes into account free-energy differences between three levels, \textsf{w,~i,~o}, representing (polar) water, (mixed) interface, and (nonpolar) octanol or lipid-hydrocarbon-tail environments, respectively.
Our modeling with these data strongly indicates that the aforementioned instability is real and in agreement with experiments \cite{AKW+10, WAR16, HSQ+16}.

For the investigation of insertion pathways we replace the three environmental levels \textsf{w,~i,~o} by an environmental field in the manner discussed in Sec.~\ref{sec:mem-env}. 
We use the density field of water (\ref{eq:1}) to specifically convert the steps $\Delta G_\mathsf{wo}^{(n)}$ from Wimley-White data \cite{WW96, WCW96, WW98, WW99, Guy85, HW11, MF11} for the residue at sequential position $n$ into fields of gradual change:
\begin{equation}\label{eq:38} 
G_\mathrm{res}^{(n)}(x)=\Delta G_\mathsf{wo}^{(n)}\big[1-\rho_\mathrm{w}(x)\big].
\end{equation}
The side-chain contribution to the free-energy of the peptide then becomes,
\begin{equation}\label{eq:39}
G_\mathrm{SC}(x_{n_\mathrm{K}},\theta_\mathrm{N},\theta_\mathrm{C})=
\sum_{n=1}^{N_\mathrm{R}} G_\mathrm{res}^{(n)}(x_n),
\end{equation}
with the dependence of $x_n$ on $x_{n_\mathrm{K}},\theta_\mathrm{N},\theta_\mathrm{C}$ from (\ref{eq:29}).
From here on we use the specifications $x_\mathrm{a}=17.8$\AA, $x_\mathrm{b}=26.3$\AA, $x_\mathrm{s}=2.0$\AA~ for the profile (\ref{eq:1}).

\begin{figure}[htb]
  \begin{center}
  \includegraphics[width=75mm]{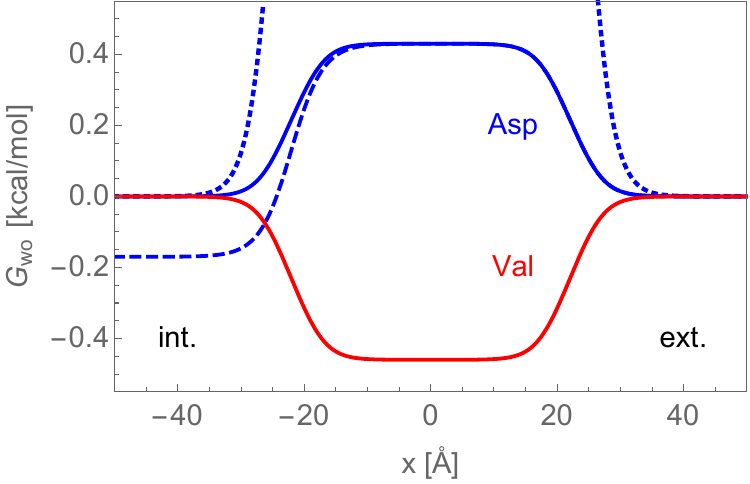}
\end{center}
\caption{Transfer-free-energy profile of one hydrophobic residue, \textsf{Val}, and one hydrophilic residue \textsf{Asp}, when protonated (solid line) or deprotonated (dashed line). The dashed line represents a hybrid profile for the same \textsf{Asp} residue, which accounts for a kinetic effect as described in the text.}
  \label{fig:f16}
\end{figure}

The transfer free energies relative to an exterior position for the hydrophobic \textsf{Val} residue and the  mildly hydrophilic \textsf{Asp} residue (when protonated) are shown in Fig.~\ref{fig:f16} (solid curves) as an illustration of this scheme. 
The deprotonated \textsf{Asp} residue is much more strongly hydrophilic, which costs significant extra free energy, as indicated by the dotted line.
The dashed line represents a modification of the Asp profile that takes into account the complication that during the insertion process, the pH is different in the exterior and interior regions.
This last point requires some explanation.

When the pH is lowered in the exterior region, all \textsf{Asp} residues quickly become protonated and, in consequence, less hydrophilic.
Their transfer-free-energy profile changes from the dotted line to the solid line in Fig.~\ref{fig:f16}.
It is well established in experiments with liposomes that the insertion process is faster than the equilibration of the pH across the membrane.
Hence the three \textsf{Asp} residues closest to the \textsf{C} terminus will only remain protonated, during the insertion process, until they come into contact with the aqueous environment in the interior region, which is still at higher pH.
The consequence in the framework of quasistatic translocations considered here is that the transfer-free energies of these \textsf{Asp} residues drop to a level below that of the exterior region, as is reflected in the dashed profile of Fig.~\ref{fig:f16}.
As the pH slowly drops inside the liposome, the modified (dashed) profile gradually turns back into the profile characterizing the protonated \textsf{Asp} (solid line).

In the projected kinetic study of peptide insertion \cite{kinpep}, this complication will, of course, be accounted for in a very different way.
The goal here is more modest.
We aim to identify the circumstances under which insertion pathways for pHLIP in the free-energy landscape exist.
There are no intrinsic time scales associated with these pathways.
However, it appears to be a necessary condition for insertion, as will be explained in Sec.~\ref{sec:inserpath}, that a differential in pH between exterior and interior regions is maintained.
In liposome experiments, this pH differential only exists for a limited time, during which insertion must be completed.

%
\section{Pathways}\label{sec:inserpath}
%
Here we pick up the thread from Sec.~\ref{sec:scen-l2}, where we designed a template for exploring adsorption and insertion pathways of a peptide consisting of two straight segments with a kink at the \textsf{Pro} position (our choice).
Any pathway taken by the peptide must be downhill in the free-energy landscape, assembled from the backbone part (\ref{eq:30}) and the side-chain part (\ref{eq:39}),
\begin{align}\label{eq:40}
G(x_{n_\mathrm{K}},\theta_\mathrm{N},\theta_\mathrm{C})
&=G_\mathrm{BB}(x_{n_\mathrm{K}},\theta_\mathrm{N},\theta_\mathrm{C})
+G_\mathrm{SC}(x_{n_\mathrm{K}},\theta_\mathrm{N},\theta_\mathrm{C})
\nonumber \\
&=\sum_{n=1}^{N_\mathrm{R}}\left[\bar{G}_\mathrm{H}(x_n)
+G_\mathrm{res}^{(n)}(x_n)\right],
\end{align}
with $x_n$ from (\ref{eq:29}).

We have adopted a very simple search procedure for descending pathways.
For each step we consider positive and negative infinitesimal increments for the three variable $x_{n_\mathrm{K}}, \theta_\mathrm{N}, \theta_\mathrm{C}$ (see Fig.~\ref{fig:f12}) and execute the step that produces the most negative $\Delta G$.
The pathway stops when none of the variable changes generates a $\Delta G<0$.
Guided by the circumstances associated with experimental investigations of pHLIP insertion into liposomes, we divide the pathways toward lower free energy into three consecutive phases.
Different environmental conditions pertain to each phase.

(A) During the \emph{adsorption} phase, the pH is high and all \textsf{Asp} residues are deprotonated.
The peptide starts out in solution close to the membrane.
Pathways with $\Delta G<0$ end with the peptide adsorbed at the interface of the membrane with the exterior aqueous region.

(B) The \emph{insertion} phase starts with the peptide in that adsorbed state but with the environment changed.
The pH is low in the exterior region and all \textsf{Asp} residues have become protonated. 
Pathways with $\Delta G<0$ end with the peptide in a trans-membrane state.
The pH remains high in the interior region.

(C) The \emph{stabilization} phase comes into play once the pH has dropped in the interior region (of a liposome), which typically happens more slowly than the insertion.
It affects the stability of the trans-membrane state.
However, pathways with $\Delta G<0$ keep the peptide inserted.

\subsection{Adsorption}\label{sec:adsorp}
\begin{figure}[htb]
  \begin{center}
\includegraphics[width=40mm]{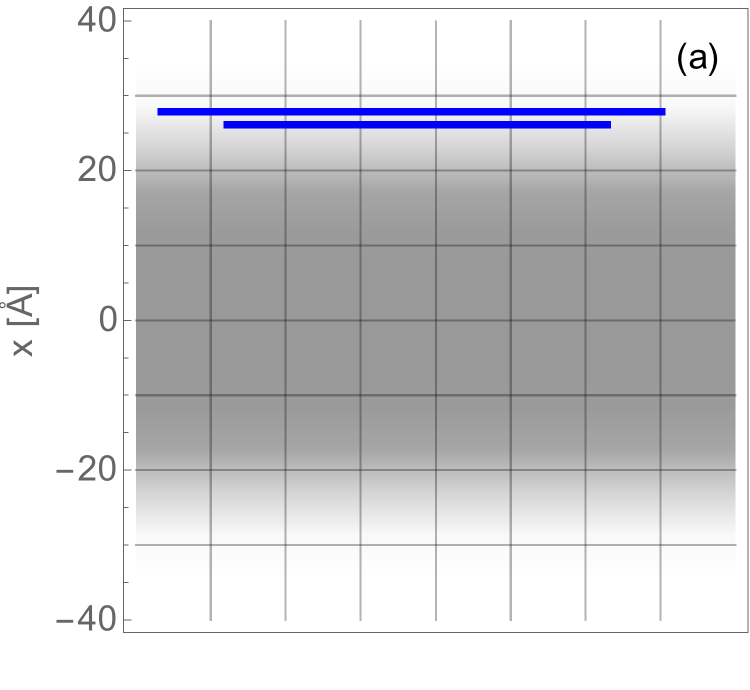}%
\hspace*{3mm}\includegraphics[width=40mm]{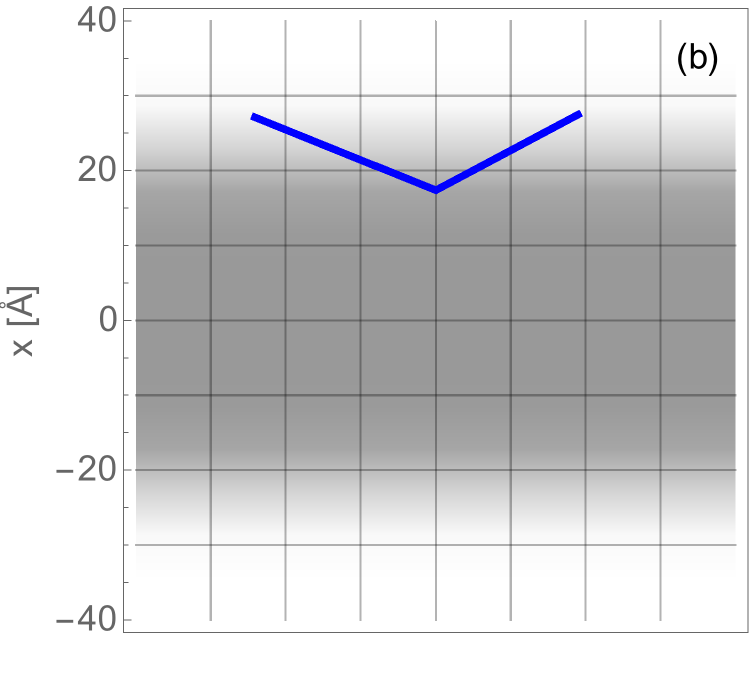}
\end{center}
\caption{Variant W6 of pHLIP modeled as two straight segments with a kink at the \textsf{Pro} position during the adsorption phase of the pathway.}
  \label{fig:f17}
\end{figure}

The gray background of Fig.~\ref{fig:f17} is a representation of the water-density profile [Fig.~\ref{fig:f3}(a)] across the heterogeneous membrane environment.
The exterior region is at the top.
At the starting point of the pathway explored here, the adsorption phase is already on its way.
The peptide is sprawled wide near the interface.

The adsorption pathway from here on consists of two legs. 
First the peptide moves toward the membrane without changing its orientation.
Along this stretch, it undergoes a conformational change from coil to mostly helix, which shortens its length.
The initial and final configurations of this first leg are shown in Fig.~\ref{fig:f17}(a).
Along the second leg of the adsorption pathway, the two segments on either side of the \textsf{Pro} kink change their orientations and the kink position continues to move toward the center of the membrane.
This brings the hydrophobic residues deeper into the nonpolar environment, yet keeps the hydrophilic residues closer to the polar environment.
The final, adsorbed equilibrium state is shown in Fig.~\ref{fig:f17}(b).

\begin{figure}[htb]
  \begin{center}
\includegraphics[width=40mm]{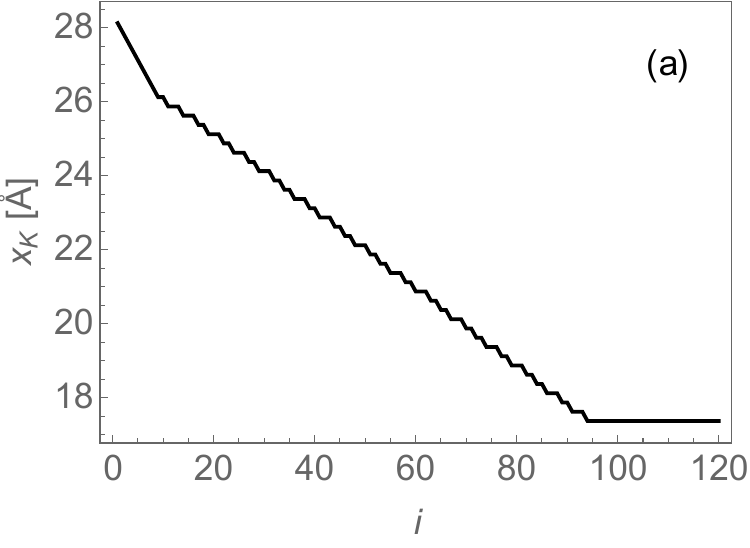}%
\hspace*{3mm}\includegraphics[width=40mm]{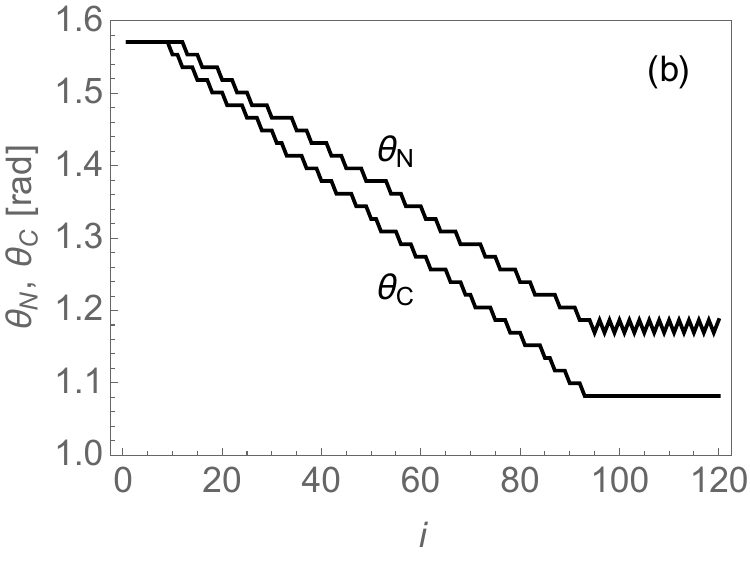}
\includegraphics[width=40mm]{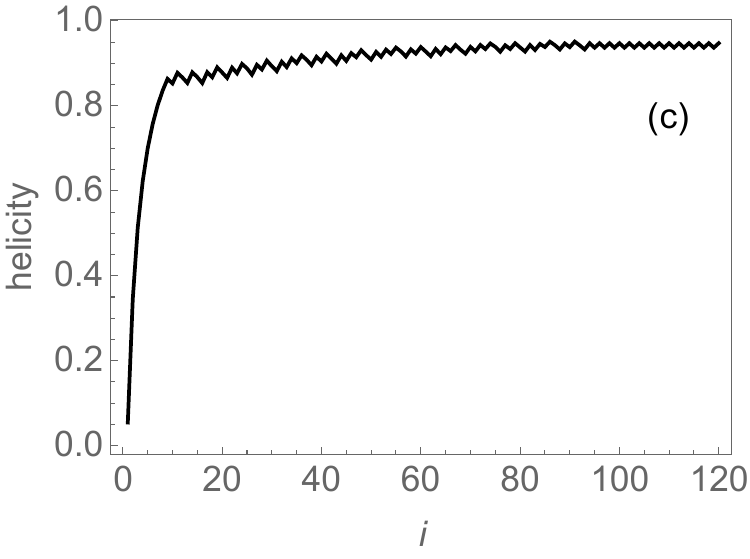}%
\hspace*{3mm}\includegraphics[width=40mm]{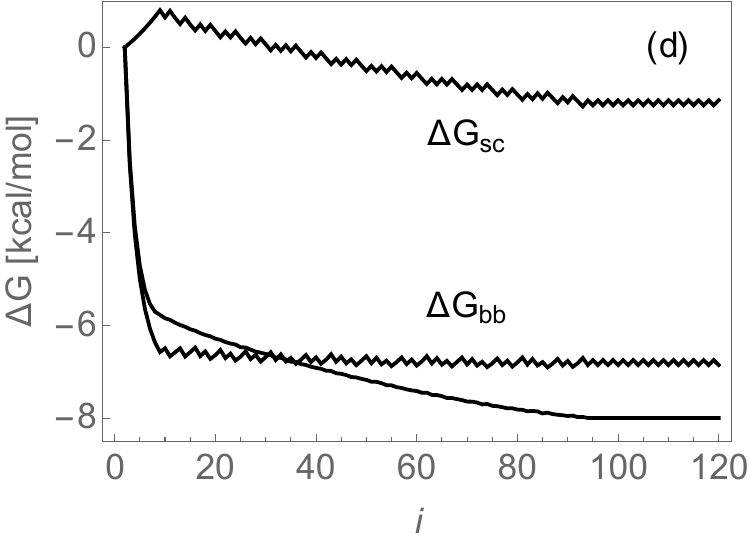}
\end{center}
\caption{(a) Kink position, (b) angles (away from vertical orientation) of the two segments, (c) helicity, (d) change in free energy (backbone, side-chain contributions and total) versus step number $i$ along the adsorption pathway.}
  \label{fig:f20}
\end{figure}

Additional information of a more quantitative nature about the adsorption pathway is compiled in Fig.~\ref{fig:f20}.
The two legs are clearly discernible in the variations of the kink position [panel (a)] and the angles of orientation of the two segments [panel (b)]. 
The first leg ends after step $i=8$.
Here the angles begin to vary and the kink moves at a slower rate, meaning down a more shallow slope.
The pathway ends after step $i=100$.
The two legs are more clearly recognizable in some data than in others.
The zigzag lines visualize the search protocol described earlier.

The conformational change from coil to helix happens almost entirely during the first leg.
Here the decrease in free energy is driven by the backbone contribution.
The conformational change, which involves the backbone, takes place during the first leg. 
An environmental differentiation restricted by conformational constraints takes place during the second leg.
Hydrophobic residues (near the center of the peptide) move further into membrane while hydrophobic residues stay closer to water. 
We have chosen the starting point of the adsorption phase such that the overall change of free energy ($\sim-6$kcal/mol) is consistent with caloric experiments \cite{RAS+08, WMT+13}.

\subsection{Insertion}\label{sec:insert}

\begin{figure}[htb]
  \begin{center}
\includegraphics[width=40mm]{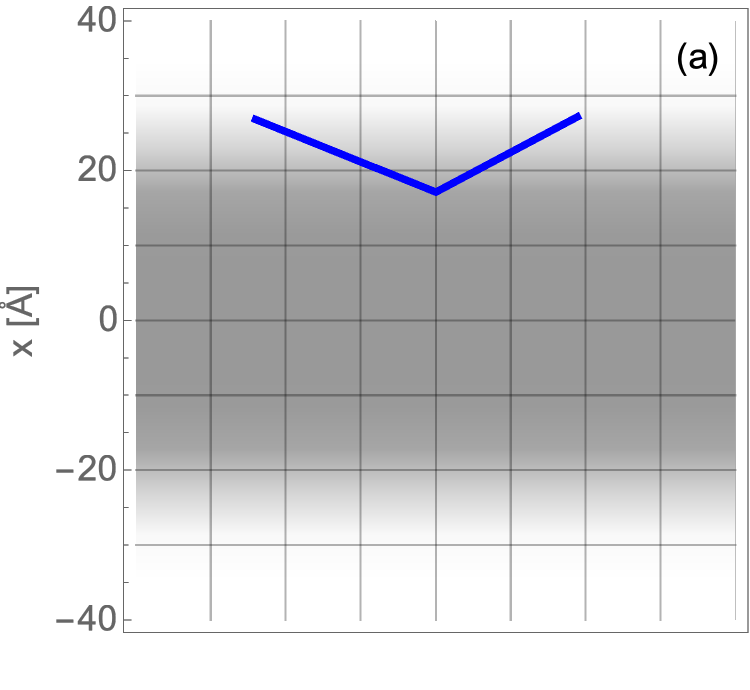}%
\hspace*{3mm}\includegraphics[width=40mm]{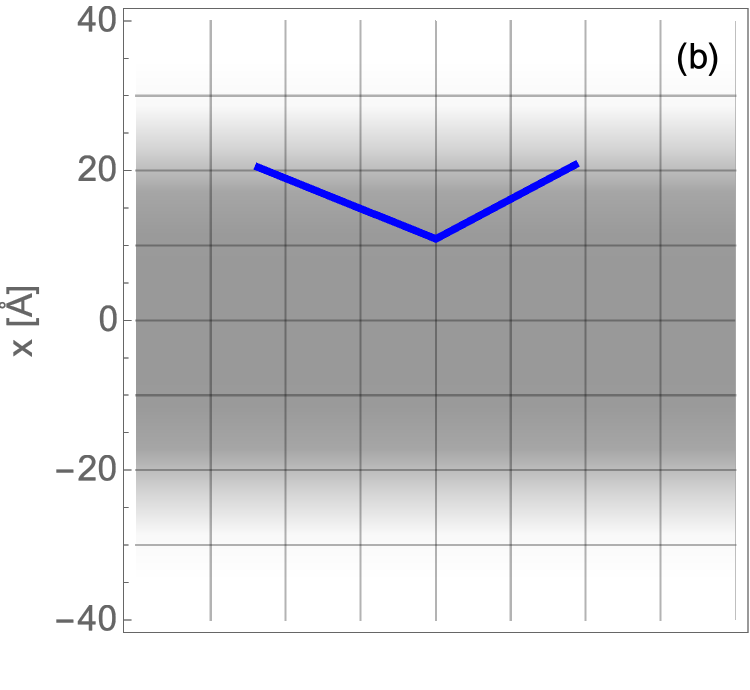}
\includegraphics[width=40mm]{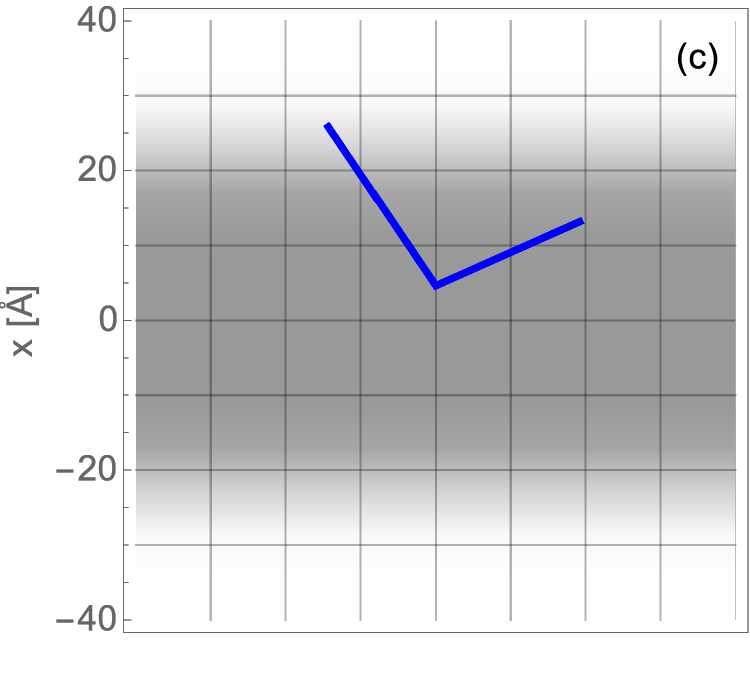}%
\hspace*{3mm}\includegraphics[width=40mm]{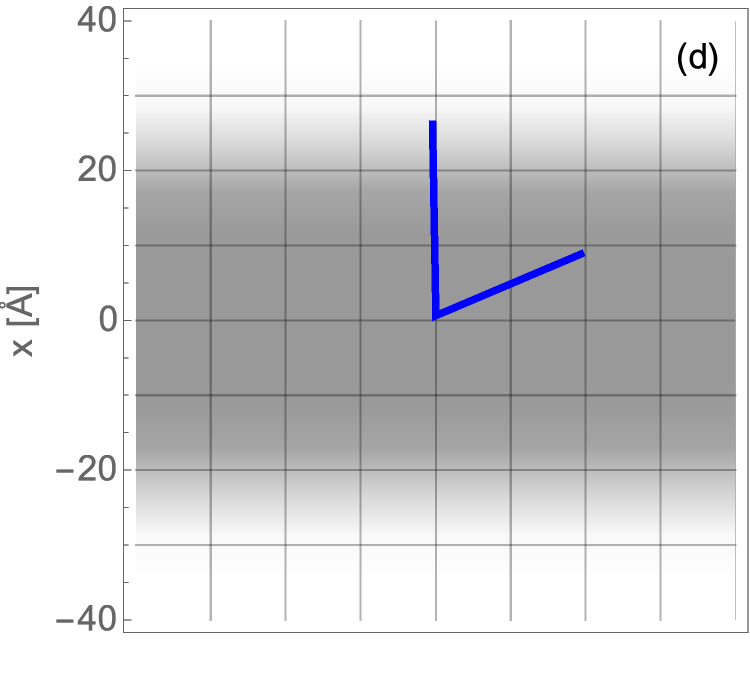}
\includegraphics[width=40mm]{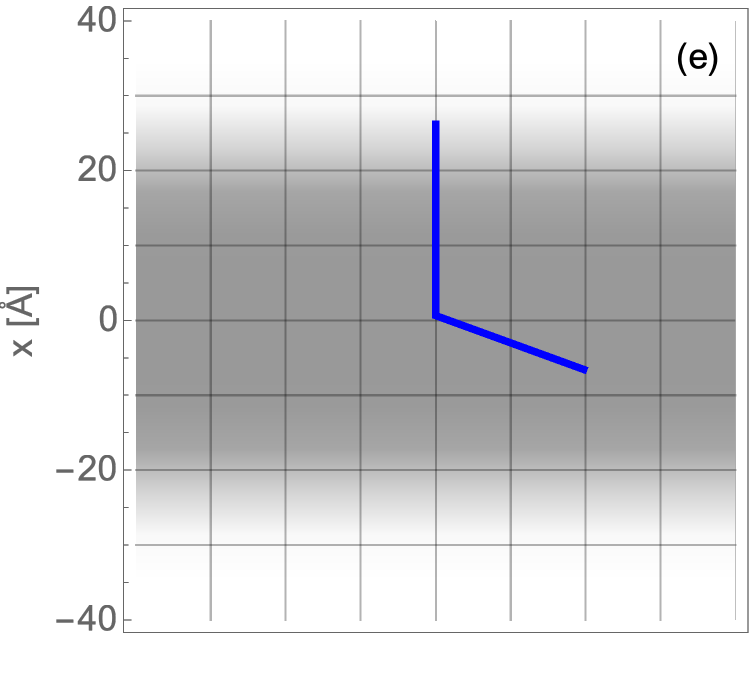}%
\hspace*{3mm}\includegraphics[width=40mm]{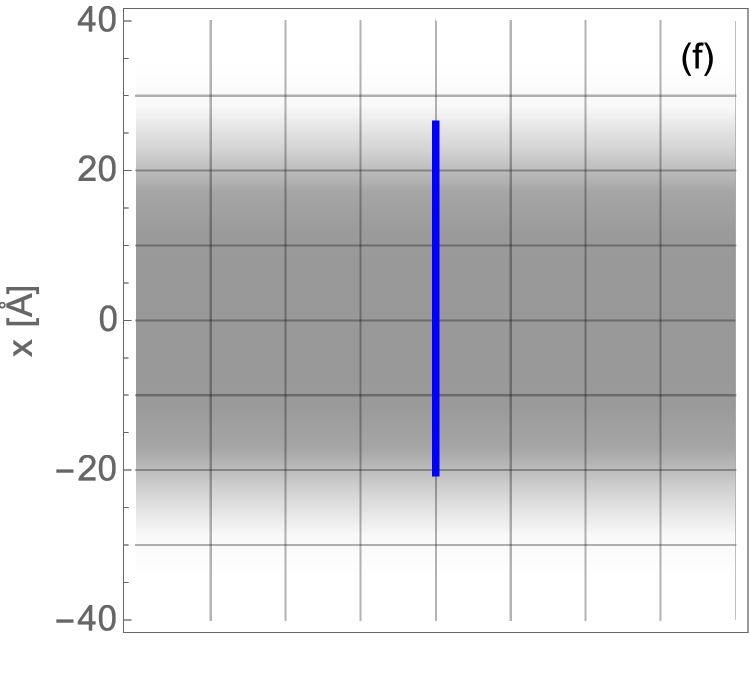}
\end{center}
\caption{Variant W6 of pHLIP modeled as two straight segments with a kink at the \textsf{Pro}18 position during the insertion phase of the pathway.}
  \label{fig:f18}
\end{figure}

When the pH is lowered in the exterior region, the final state of the adsorption pathway becomes unstable.
It becomes the initial state of the next phase of the pathway [Fig.~\ref{fig:f18}(a)].
All \textsf{Asp} residues are now protonated \cite{ADS+07, MKA+10}.
The destabilizing agent is the switch of their transfer-free-energy profile from the deprotonated to the hybrid version.
This modification opens up pathways of descending free energy, which we explore using the previously described protocol.
What emerges turns out to be an insertion pathway.
It consists of three legs, as visualized in Fig.~\ref{fig:f18}.

Along the first leg [panels (a)-(b)], the peptide sinks somewhat deeper into the membrane with no significant change in the orientation of the two segments on either side of the kink at the \textsf{Pro} position.
The protonation has made the \textsf{Asp} residues significantly less hydrophilic.
This has weakened their resistance against the pull of the hydrophobic residues into the nonpolar environment.
Hence the downward translocation of the entire peptide.

Along the second leg [panels (b)-(d)], the segment on the left with the \textsf{N} terminus at its end straightens up while the other segment does not rotate significantly.
The \textsf{Pro} kink position moves further toward the center of the membrane.
Protonation has taken place primarily near the \textsf{C} terminus. 
The positive charges at the \textsf{N} terminus and at the \textsf{Arg} residue on the same segment are still present. 
Reorienting that segment into a trans-membrane direction keeps those charges in or near the polar environment and allows the hydrophobic residues to move deeper into nonpolar environment.

The third leg of the insertion pathway [panels (d)-(f)] is mainly driven by the forces acting on the segment with the \textsf{C} terminus at its end.
There are competing forces in action. 
The hydrophobic forces lead the initial descent in free energy toward a horizontal orientation of that segment.
The subsequent descent in free energy toward the trans-membrane orientation is guided by the forces acting on the protonatable \textsf{Asp} residues and the \textsf{C} terminus.
Recall that the interior region is still at high pH, which significantly enhances the pull of the protonatable contacts toward the nonpolar environment, where deprotonation takes place.

\begin{figure}[b]
  \begin{center}
\includegraphics[width=40mm]{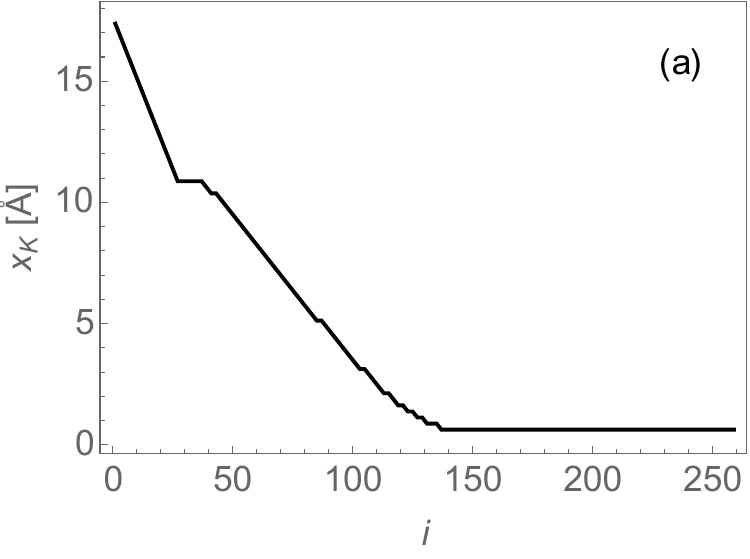}%
\hspace*{3mm}\includegraphics[width=40mm]{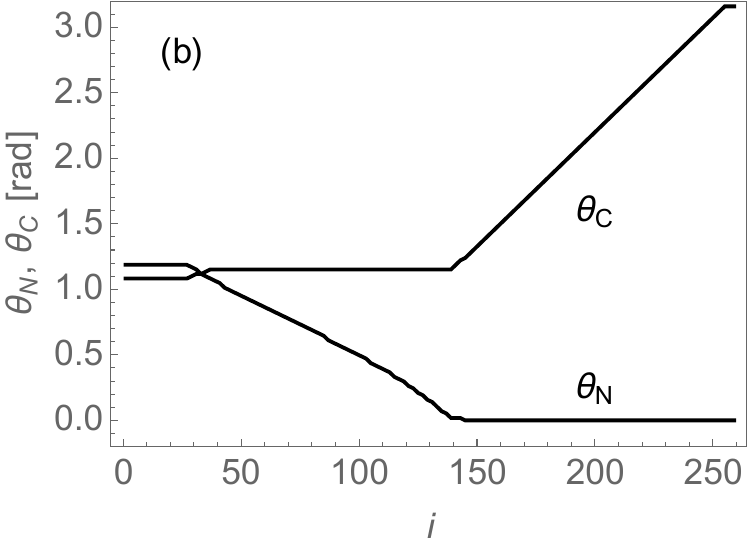}
\includegraphics[width=40mm]{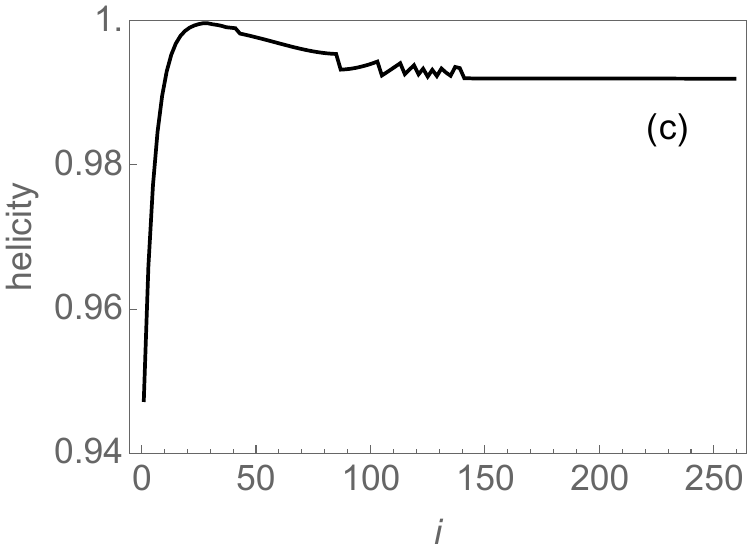}%
\hspace*{3mm}\includegraphics[width=40mm]{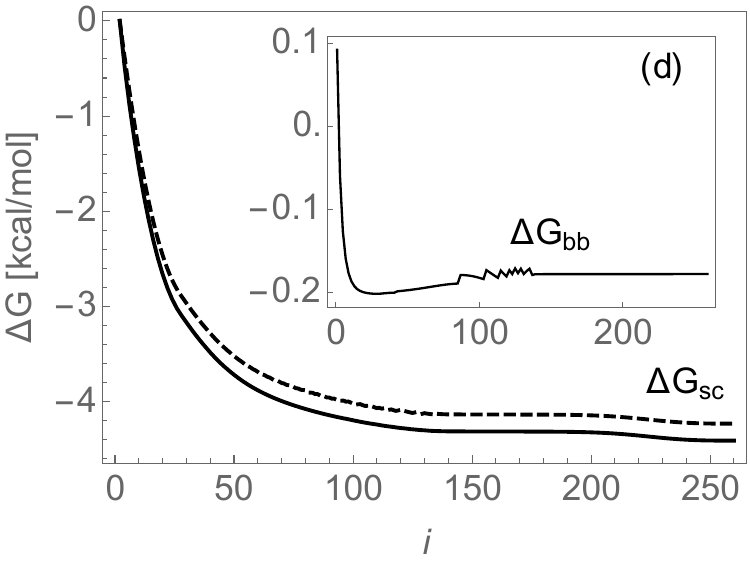}
\end{center}
\caption{(a) Kink position, (b) angles (away from vertical orientation) of the two segments, (c) helicity, (d) change in free energy (backbone, side-chain contributions and total) along the insertion pathway.}
  \label{fig:f21}
\end{figure}

Figure~\ref{fig:f21} provides additional, more quantitative information about the insertion pathway. 
The three legs are most clearly discernible in the kink position and the two angles of orientation.
The reorientation of the two segments during the second and third legs is almost completely sequential: first the \textsf{N} terminus backs out into the exterior region while the kink position moves toward the center; then the \textsf{C} terminus moves across into the interior region while the kink position remains stationary.

The helicity (already strong) reaches near saturation during the first leg, but then decreases somewhat as the \textsf{N} terminus reestablishes closer contact with the exterior aqueous region.
The change in free energy during insertion originates predominantly from the side chains.
The drop of backbone free energy is less than 0.3 kcal/mol and happens quickly while the helicity increases. 
The drop in side-chain free energy is, for the most part, spread across the first two legs of the insertion pathway. 
Its amount of $\sim 4$kcal/mol is a bit higher than what caloric experiments predict, but not by much \cite{RAS+08, WMT+13}.

In our study we have worked with a single density field of water, not taking into account any possible variations due to a change in pH. 
We did confirm though that the insertion pathways described above are robust under small changes of the parameters $x_\mathrm{a}$, $x_\mathrm{b}$.
Variations up to 5\% in either parameter did not produce any qualitative changes in the insertion pathway and the stability of the transmembrane state. 
The only proviso is that for smaller $x_\mathrm{a}$, the initial position of the peptide must be closer to the membrane for adsorption to ensue.

\subsection{Stability}\label{sec:protleak}
In an experiment that uses liposomes instead of biological cells, the low level of pH imposed on the exterior region will slowly leak into the interior region.
This happens on a significantly slower time scale than the insertion process.
There is clear experimental evidence that pHLIP stays inserted as the pH equilibrates at a low level.
Hence the third phase of the pathway must ensure that when we change the \textsf{Asp} transfer-free-energy profiles from the hybrid version to the protonated version the trans-membrane state remains stable \cite{RSA+07, KWW+12}.

\begin{figure}[htb]
  \begin{center}
\includegraphics[width=40mm]{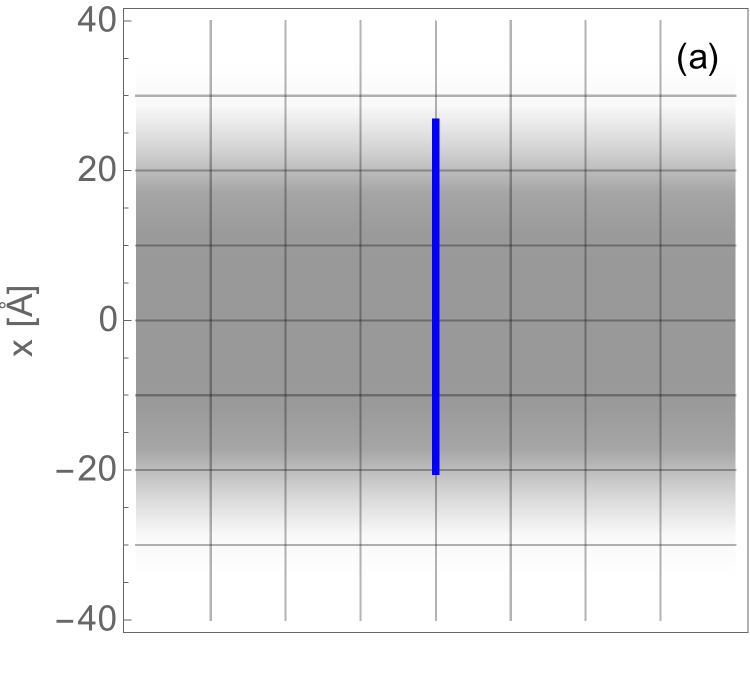}%
\hspace*{3mm}\includegraphics[width=40mm]{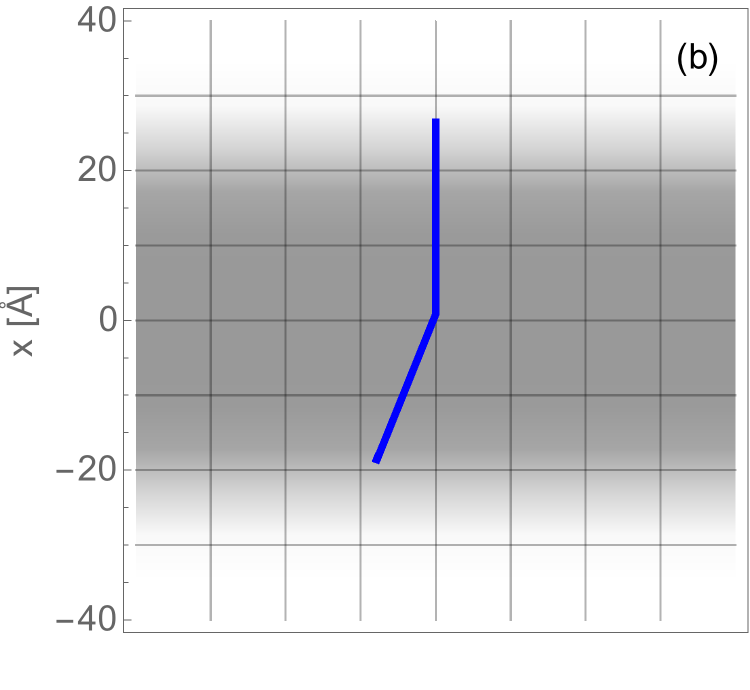}
\end{center}
\caption{Variant W6 of pHLIP modeled as two straight segments with a kink at the \textsf{Pro}18 position during the third phase of the pathway.}
  \label{fig:f19}
\end{figure}

This turns out to be the case indeed as illustrated in Fig.~\ref{fig:f19}. 
Panel (a) shows the final state of the second phase and panel (b) the final state of the third phase.
The low pH in the interior region reprotonates the \textsf{C} terminus and the \textsf{Asp} residues near it.
Hence their counteraction against the hydrophobic forces on the \textsf{Leu} residues (also near the \textsf{C} terminus) weakens.
This has the effect that the segment with the \textsf{C} terminus sticking into the interior region changes its orientation somewhat to find the new local free-energy minimum.

The kink position and the orientation of the segment with the \textsf{N} terminus at its end do not undergo significant changes.
The descent in free energy is small in comparison to the previous two stages.
It amounts to less than 10cal/mol.
The important message is that the peptide stays in a trans-membrane state.
Note that insertion (second stage) only happens, according to the protocol of our pathway exploration, if the pH remains high in the interior region until the trans-membrane state has been realized.
However, once the trans-membrane state is realized it remains stable even after the pH has dropped in the interior region. 

Let us emphasize that Figs.~\ref{fig:f20} and \ref{fig:f21} are not meant to suggest any times scales for the adsorption and insertion processes, respectively. 
Our protocol for the exploration of free-energy landscapes merely searches for downhill directions and then takes steps accordingly with no attempts undertaken to optimize directions for steepest descent. 
A kinetic study of the pHLIP adsorption and insertion processes is an entirely different project \cite{kinpep}.

%
\section{Conclusion and outlook}\label{sec:con-out}
%
Processes of peptide insertion into a membrane are highly complex in their kinetics and can be very diverse in their energetics.
In this work an effort has been described that focuses on the energetics of a particular scenario.
The specific goal has been to demonstrate plausible insertion pathways in a free-energy landscape assembled from contributions identified as dominant.
We built that landscape from enthalpic and entropic contributions associated with internal and external H-bonds along the peptide backbone and from transfer free energies of side chains between hydrophilic and hydrophobic regions of the membrane environment.

The advantage of the approach centered on energetics is that the problem associated with the inevitably broad range of time scales has come into play only marginally.
The investigation of free-energy landscapes and the search for insertion pathways is not meant to be a substitute for a molecular dynamics simulation or other form of kinetic study \cite{DQL13, DFST16, GV09, VSS+18, MBT08, MS92, MS93, IB05, NWG05, UU08, USU10, UDK+10}.
It is an exploration of the circumstances under which kinetic processes associated with peptide insertion are likely to take place. 

Insertion pathways as explored in this work are characteristic of quasistatic processes. 
The free-energy landscapes provide road signs, but no time tables.
The road signs thus established are not infallible because the free-energy landscapes in use are manifestly incomplete. 
A strong case can be made that what has been left out is more productively taken into account in a kinetic study or a molecular dynamics simulation.
The most relevant missing pieces represent different aspects of lipid-peptide interactions.
Some are predominantly enthalpic and other predominantly entropic in nature.

(i) The lateral pressure profile of the membrane, which provides mechanical stability to the bilayer structure against perturbations of various kinds, is known to have a characteristic shape with regions of positive deviations from ambient pressure sandwiching a narrow band of negative deviation just inside the lipid headgroups.
These empirically established pressure variations are significant \cite{Mars96, Cant97, Cant99, GS04, GBM06, Mars07, BMT10, ZL13, DFST16, DM16}.
 
(ii) In the (trans-membrane) state III the predominently hydrophobic and $\alpha$-helical center of pHLIP is flanked by coil segments that are more hydrophilic.
In a positive (negative) mismatch, the hydrophobic center is too long (short) to be comfortably accommodated in the membrane.
Each type of mismatch elicits a distinct response from both the peptide and the membrane.
In a positive mismatch, the membrane tends to increase its width locally.
Sheltering additional hydrophobic residues from water costs elastic energy.
That cost can be lowered if the helical axis tilts its trans-membrane orientation away from the bilayer normal.
To a negative mismatch the membrane responds with a local thinning,
which puts the helix under tension.
Some of that tension may be released by a partial change from $\alpha$-helix to the more tightly wound $3_{10}$-helix \cite{BK92, Mill95, Kill98, PK03, SEUS12, DA13, note15}.

(iii) The lipid molecules in the bilayer, typically bent into a liposome of spherical shape, are in a macrostate which represents a two-dimensional fluid.
The positional ordering of the headgroups is then of short range over some coherence length $\xi$.
The undisturbed lipid bilayer is characterized by a uniform entropy density with two contributions, one representing positional disorder of the headgroups and the other orientational and conformational disorder of the hydrocarbon tails \cite{BBH96, CKT05}.
Headgroups and tails perform a free-energy balancing act of sorts.
An increase in bilayer width produces more space for the tails to explore and requires work by the headgroups against the pressure of the surrounding water, thus implying $T\Delta S>0$ and $\Delta H>0$, respectively.
A decrease in the average number of headgroups per unit area increases their positional freedom, implying $T\Delta S>0$, but compresses the tails, implying $\Delta H>0$.
Balance is reached when $\Delta G=\Delta H-T\Delta S=0$.

(iv) The presence of pHLIP produces a contact line with lipid molecules.
Along that contact line, the aforementioned thermal equilibrium of headgroups and tails is being disturbed \cite{LHH95, WHLH95}.
The dominant changes involve an entropy reduction in both headgroups and tails.
The former tend to line up tightly against the foreign object and form an ordered layer of width roughly equal to the aforementioned coherence length $\xi$. 
pHLIP going into the (adsorbed) state II throws the normal pressure out of balance.
Fewer tails must exert the same force per unit area.
They can do that only under higher compression, facilitated by membrane thinning as noted before.
This, in turn, leads to an entropy reduction.
When pHLIP inserts the contact line becomes much shorter.
The associated decrease in $\Delta G$ dominated by a positive $T\Delta S$ is a factor favoring insertion thermodynamically.
J\"ahnig \cite{Jaeh83} investigated this effect under the name \emph{lipohobic effect} and used a coherence length of $\xi\simeq15$\AA.

(v) What we have been calling water-lipid interface involves, in fact, an electrolyte with variable ion concentration on one side.
In our study we have taken into account one particular aspect of this presence and variation, namely the effect of the pH on the protonation status of negatively charged residues.
There are different, well documented ways in which ion content of the water affects the lipid bilayers, some of which overlap with the previous items of this list \cite{ASR16, KFD16, Frie18, Dobr19, DWM+19, DTCG20, HSD21, XHL+22}.

Taking into account effects (i)-(v) calls for a more detailed model of the peptide than is being used in this study.
The stochastic modeling intended to be used for pHLIP insertion kinetics  \cite{kinpep} is a Markov chain model, which promises interesting points of comparison with molecular dynamics simulations. 
One strong point of Markov chain modeling is that it can deal with a range of different time scales in a most transparent and efficient way.
The price to be paid for that advantage is that interactions on a microscopic scale are accounted for more summarily than a simulation does at the molecular dynamics level.

\acknowledgments
We are grateful to Yana K. Reshetnyak and Oleg A. Andreev for suggesting this study and for generously offering much help along the way.

\appendix

\section{Statistically interacting polymer links}\label{sec:appa}
The origin in quantum many-body theory \cite{Hald91a, Wu94, Isak94, Anghel, PMK07} of the methodology used here and its adaptation to problems of current interest in classical statistical mechanics \cite{LVP+08, copic, picnnn, pichs, GKLM13, sivp, janac2} has already been presented from several different angles. 
Among these works are papers dedicated to polypeptides \cite{cohetra} and to double-stranded DNA \cite{mct1}.

Here we summarize those results from Ref.~\cite{cohetra} which are being used (first in Sec.~\ref{sec:prof}) as the main building blocks for extensions reported in this work.
The microscopic model for the coil-helix transition of a long polypeptide at a water-lipid interface solved in \cite{cohetra} has three parameters: the growth parameter $t$, the nucleation parameter $\tau$, both continuous, and the (discrete) range parameter $\mu$, which numbers the coil states available to each residue (see Fig.~\ref{fig:f11}). 
Here we consider the cases $\mu=2$ and $\mu=\infty$ at $\tau>0$ and use selected results.
We quote the relevant expressions in ways easy to trace back to their derivations.

The scaled Gibbs free energy in closed form reads
\begin{equation}\label{eq:a1} 
 \frac{\bar{G}(t,\tau)}{k_\mathrm{B}T} =-\ln\left(\frac{1+w(t,\tau)}{t}\right),
\end{equation}
where, for $\mu=2$,
\begin{align}\label{eq:a2} 
&w(t,\tau)=\frac{1}{3}\left[x+2\sqrt{x^2+3y}\cos\frac{\varphi}{3}\right], \\
&\tan\varphi=\frac{\sqrt{27(4y^3+y^2x^2+18yx^2+4x^4-27x^2)}}
{x(2x^2+9y-27)},\nonumber
\end{align}
with $x\doteq t-1$, $y\doteq 1+t\tau$ and $0\leq\varphi<\pi$, and, for $\mu=\infty$,
\begin{equation}\label{eq:a3}
w(t,\tau)=\left\{ \begin{array}{ll} 2 & : 0\leq t\leq t_\mathrm{c}, \\ 
t-1+{\displaystyle \frac{t\tau}{\lambda}} & : t>t_\mathrm{c},
\end{array} \right.
\end{equation}
with
\begin{equation}\label{eq:a4} 
\lambda(t,\tau)\doteq \frac{1}{2}\left[t-1+\sqrt{(t+1)(t-3)+4t\tau}\,\right]
\end{equation}
and 
\begin{equation}\label{eq:a5} 
t_\mathrm{c}\doteq \frac{3}{1+\tau}.
\end{equation}
The expressions for helicity (order parameter) and entropy, inferred from first derivatives of $\bar{G}$ become
\begin{equation}\label{eq:a6}
\bar{N}_\mathrm{hl}(t,\tau) = 
  \frac{t}{1+w}\frac{\partial w}{\partial t},
\end{equation}
\begin{align}\label{eq:a7} 
 \frac{\bar{S}(t,\tau)}{k_\mathrm{B}} &=
 \ln(1+w) - \frac{1}{1+w}  
 \left[t\ln t\frac{\partial w}{\partial t} +
   \tau\ln\tau\frac{\partial w}{\partial \tau}\right],
\end{align}
respectively.
They are too unwieldy for $\mu=2$ to be reproduced here but fairly concise for $\mu=\infty$:
\begin{equation}\label{eq:a8} 
 N_{\mathrm{hl}}(t,\tau)  = 
  \begin{cases}\displaystyle
    0
    &: 0\leq t < t_{c},
    \\ \displaystyle
    1-\frac{t\tau}{\lambda^{2}-1+t\tau} &: t \geq t_{c},
  \end{cases}
\end{equation}
\begin{align}\label{eq:a9} 
  \frac{\bar{S}(t,\tau)}{k_\mathrm{B}} &=\ln\left(1+\frac{\tau}{\lambda}\right) \nonumber \\ &\hspace{5mm}+
  \frac{t\tau}{\lambda^{2}-1+t\tau}
  \left(\ln t
    -\frac{\lambda^{2}-1}{t(\lambda+\tau)}\ln\tau
    \right).
\end{align}
Finally the scaled enthalpy, 
\begin{align}\label{eq:a10} 
\frac{\bar{H}(t,\tau)}{k_\mathrm{B}T} &=\frac{\bar{G}(t,\tau)}{k_\mathrm{B}T}+\frac{\bar{S}(t,\tau)}{k_\mathrm{B}}
\nonumber \\ &\hspace{0mm}
=     \frac{t\tau}{\lambda^{2}-1+t\tau}
    \left(\ln t
    -\frac{\lambda^{2}-1}{t(\lambda+\tau)}\ln\tau
    \right),
\end{align}
is of importance in this work.

\section{Backbone effects}\label{sec:appc}
It is instructive to take a look at the backbone contributions to the landscapes of helicity and free energy for the case of fixed angles $\theta_\mathrm{N}=0$, $\theta_\mathrm{C}=\pi$, notwithstanding its limitations.
The simplest case replaces the second Eq.~(\ref{eq:16}) by 
\begin{equation}\label{eq:10} 
l(x)=l_\mathrm{h}=1.5\mathrm{\AA}, 
\end{equation}
implying that the position coordinate of successive residues progresses uniformly and independently of conformation.
The value chosen in (\ref{eq:10}) is accurate for a helical segment but shorter than most coil segments.

\begin{figure}[htb]
  \begin{center}
\includegraphics[width=40mm]{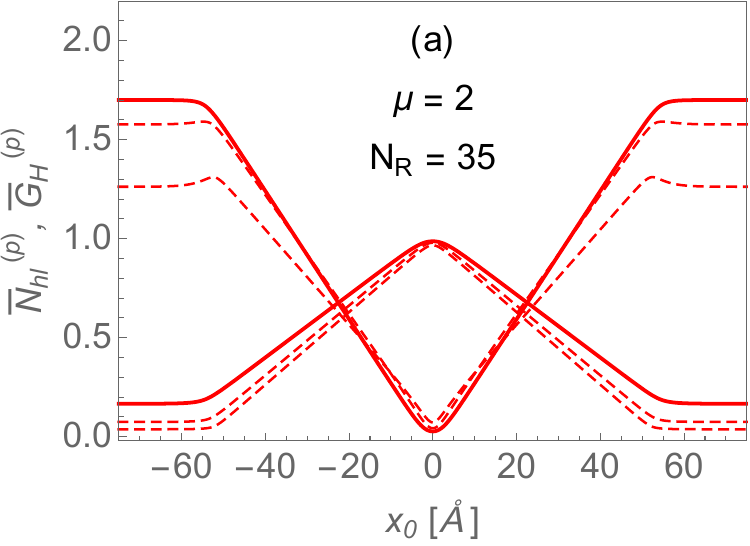}\hspace*{3mm}\includegraphics[width=40mm]{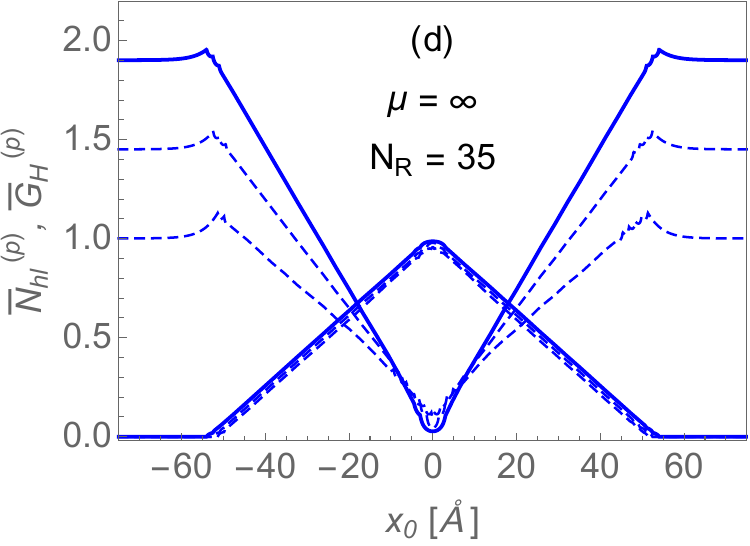}
\includegraphics[width=40mm]{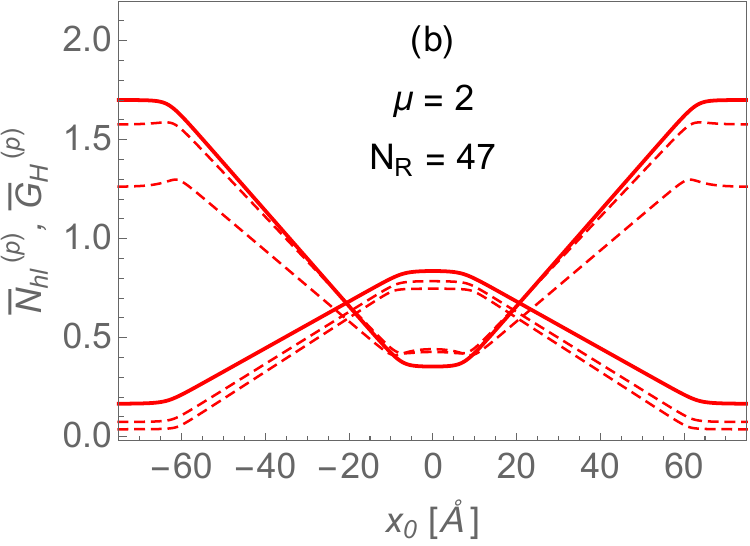}\hspace*{3mm}\includegraphics[width=40mm]{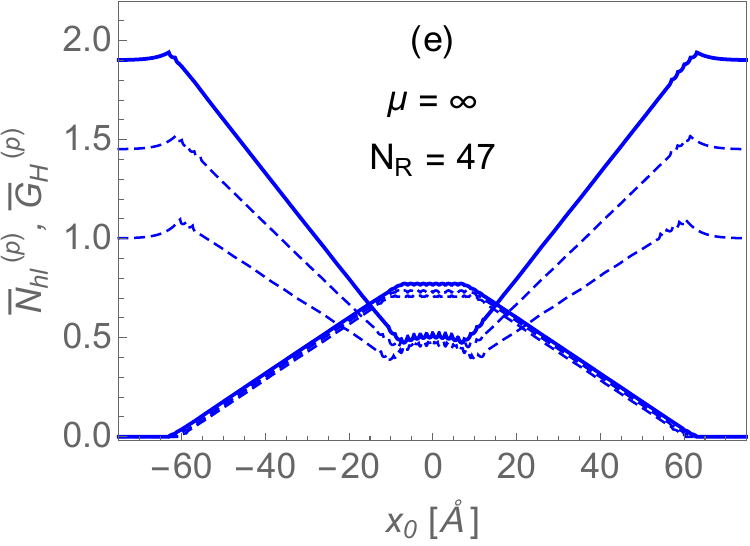}
\includegraphics[width=40mm]{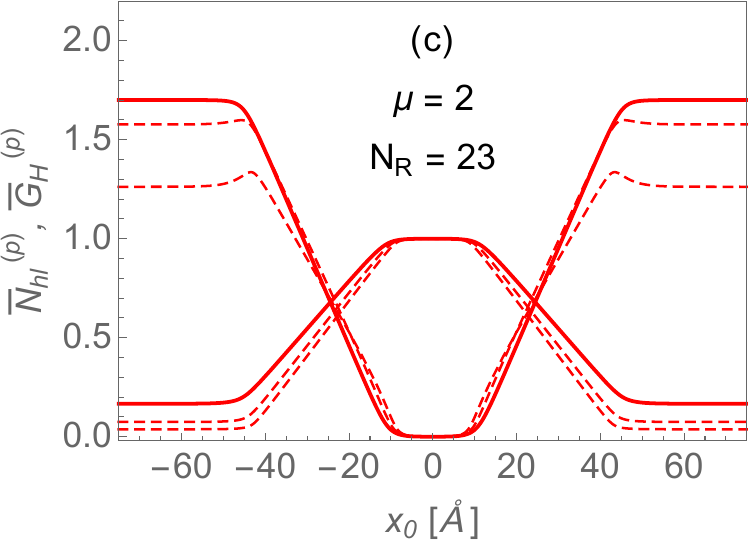}\hspace*{3mm}\includegraphics[width=40mm]{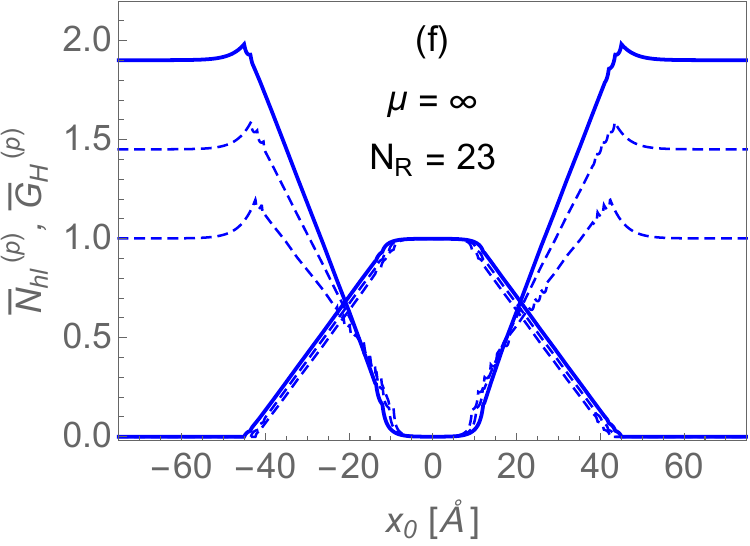}
\end{center}
\caption{Helicity (peaked at center) and scaled free-energy of peptide with $N_\mathrm{R}$ residues oriented and positioned as described in the text versus the coordinate $x_0$ of the central residue for $\mu=2$ (left) and $\mu=\infty$ (right) and three sizes. The solid and dashed curves pertain to the values $\alpha_\mathrm{H}=1$ and $\alpha_\mathrm{H}=1.05,1.1$ of the enthalpy parameter, respectively.}
  \label{fig:f8}
\end{figure}

Free-energy and helicity landscapes thus predicted are shown in Fig.~\ref{fig:f8} for $\mu=2,\infty$ and $N_\mathrm{R}=35,47,23$.
The first row of landscapes $(N_\mathrm{R}=35)$ is tailored to represent some variant of pHLIP, the second row a significantly longer peptide $(N_\mathrm{R}=47)$, and the third row a significantly shorter peptide $(N_\mathrm{R}=23)$.
We only consider enthalpy parameters that are neutral $\alpha_\mathrm{H}=1$ or represent a gain $\alpha_\mathrm{H}=1.05, 1.1$.

The case $\mu=\infty$ assigns more entropy to coil segments than the case $\mu=2$. 
The effect on the results is significant, but produces no qualitative changes.
The helicity landscapes are almost independent of the enthalpy parameter.
That parameter affects the free-energy primarily in the aqueous environment as expected.

Insertion into the membrane is clearly favored in all three cases and for both variants of the model.
The plots also tell us that insertion is accompanied by a conformational change from coil to helix.
For the longest peptide the minimum in free energy is not as deep and the maximum in helicity is not as high as is the case for the two shorter ones.
The obvious reason is that the former has significant flanking ends that remain in water.

Of particular interest is the free energy barrier that separates states with the center of the peptide in aqueous or membrane environments, the former mostly in coil conformation and the latter in helix conformation. 
This free-energy barrier is very shallow for $\mu=2$ and only exists if $\alpha_\mathrm{H}>1$.
For $\mu=\infty$, on the other hand, it is more conspicuous and present even for $\alpha=1$. 
This difference is related to the higher entropy that coil segments must shed if $\mu=\infty$ when they order into helix segments before they can cash in the enthalpic benefit of the lipid environment.

One message we take from this simplest case is that insertion is not automatic.
An environmental change may be needed to push the peptide over the barrier.
An increase in acidity is known to do the trick.
It neutralizes negatively charged residues (e.g. \textsf{Asp} and \textsf{Glu}) via protonation.
The consequences for pHLIP are well documented by experiments \cite{RSA+07, RAS+08, WMT+13, ADS+07}.
Let us recall that the extent of insertion as predicted by free-energy landscapes and the extent of ordering as predicted by helicity landscapes can be directly monitored experimentally, namely by \textsf{Trp} fluorescence and by circular dichroism experiments, respectively \cite{KWW+12,TYM+09, KWW+12, VSS+18}.

An improved level of modeling takes into account that the distance between adjacent residues is different in the coil and helix conformations as reflected in Eqs.~(\ref{eq:16}).
The distance between successive residues now depends on the local conformation of the backbone at their position in the membrane environment.
The main features in the results of the improved model (not shown) remain qualitatively the same.


\section{Environmental levels}\label{sec:appd}

In a first round of estimates we may consider each residue placed in one of three distinct environments: polar (\textsf{w}), interface (\textsf{i}), or nonpolar (\textsf{o}).
For the transfer free energies between any two environments we use the Wimley-White interface and octanol scales from \cite{WW96, WCW96, WW98, WW99}. 
State I (see Sec.~\ref{sec:scen-l2}) has all residues in solution. 
For state II we assume that all residues are at the interface. 
With that assumption we use $\Delta G_\mathrm{wi}$ for all residues to calculate the transfer free energy of the entire peptide. 
In this way we get one peptide transfer free energy,
\begin{equation}\label{eq:32} 
 \Delta G_\mathrm{I-II}^\mathrm{hpH}=+0.18\mathrm{kcal/mol},
\end{equation}
if the water is at high pH and another peptide transfer free energy,
\begin{equation}\label{eq:33} 
\Delta G_\mathrm{I-II}^\mathrm{lpH}=-7.62\mathrm{kcal/mol}, 
\end{equation}
if the water is at low pH. 
In the former we assume that the negatively charged residues are deprotonated and in the latter we assume that they are protonated.

The more important result is (\ref{eq:32}). 
It is well-established that at high pH pHLIP coexists in states I and II. 
Our rough estimate, which favors state I but only slightly, by a small fraction of $k_\mathrm{B}T$, is consistent with empirical evidence.
A refined model will take into account that adsorbed pHLIP is largely in coil conformation. 
The mechanical flexibility of this conformation allows the hydrophobic residues to be closer to the interface and some hydrophilic ones to stick out into water.
This will lower the free-energy prediction for state II relative to state I sufficiently to make $\Delta G_\mathrm{I-II}^\mathrm{hpH}$ negative but not nearly as much as (\ref{eq:33}).
A slightly negative value of $\Delta G_\mathrm{I-II}^\mathrm{hpH}$ is most consistent with experimental evidence.

Next we investigate, again by rough estimate, the peptide transfer free energies 
$\Delta G_\mathrm{II-III}$ associated with insertion.
However, during the insertion process, from state II to state III, not all residues switch environment.
We assume that in state III the trans-membrane part of the peptide is in $\alpha$-helix conformation. 
The relevant bilayer width, $\sim 35$\AA, then accommodates 23 residues in an $\alpha$-helix conformation.
These residues experience the transfer \textsf{io}. 
The five residues closest to the \textsf{N} terminus and the four residues closest to the \textsf{C} terminus are assumed to remain in the interface region.
We thus obtain the following results for the variants W6 and W17 (see Fig.~\ref{fig:f9}):
\begin{align}\label{eq:34} 
& \Delta G_\mathrm{II-III}^\mathrm{hpH}=-0.5\mathrm{kcal/mol}\quad :~ \mathrm{W6},
\\ \label{eq:35} 
& \Delta G_\mathrm{II-III}^\mathrm{lpH}=-4.32\mathrm{kcal/mol}\quad :~ \mathrm{W17}, 
\end{align}
at high and low pH, respectively. 
The numbers are slightly lower for variant W30.
The more important result is (\ref{eq:35}). 
At low pH, the experimental evidence shows that insertion is clearly favored over the adsorbed state.
We also see that the transfer free energy of insertion due to hydrophobic forces associated with side chains is considerably higher than the corresponding backbone transfer free energy investigated in Sec.~\ref{sec:land}.
The result (\ref{eq:34}) is significant for the pHLIP exit process (to be discussed elsewhere).



\begin{thebibliography}{100}
\bibitem{note2}
Experiments that use a variant with the \textsf{Pro} residue (near center) and the \textsf{Ile} residue (near \textsf{N} terminus) switched as a control sample will shed light on this issue.
\bibitem{note5}
We use the name `link' for the unit of peptide backbone between successive peptide bonds as used in the mathematical modelling of Ref.~\cite{cohetra}.
\bibitem{note6}
This estimate compares the latent heat of sublimation, $L_\mathrm{sg}=680$cal/g to the latent heat of melting, $L_{sl}=80$cal/g. If both latent heats are dominated by the breaking of H-bonds, then the fraction of unsatisfied H-bonds in the liquid phase beomes $80/680\simeq0.12$. That percentage is even lower if we assume that the average bond strength is weaker in the liquid phase.
\bibitem{note8}
We keep setting $\epsilon_\mathrm{h}=2\epsilon_\mathrm{t}$ for the activation energies of hybrids ad tags, respectively as in Ref.~\cite{cohetra} for the same reason.
\bibitem{note9}
It is fair to argue that the coil becomes more constrained in the lipid environment than in the aqueous environment, which suggest that we should use a larger value of $\mu$ in the latter than in the former. To implement the restricted space available to the polypeptide coil in this way is technically difficult. It would entail approximations of little control and obscure other environmental effects. We, therefore, opt to account for the space restrictions imposed by the lipids in different ways.
\bibitem{note10}
More precisely, the deviation of enthalpy density from its chosen reference value.
\bibitem{note15}
The energetics of the competing $\alpha$ and $3_{10}$ helix conformations is a topic well within the reach of the methodology employed here as a natural extension of Ref.~\cite{cohetra} on polypeptides and Ref.~\cite{mct1} on DNA under tension.
\end{thebibliography}
\end{document}